\documentclass[sigconf]{acmart}

\AtBeginDocument{%
  \providecommand\BibTeX{{%
    \normalfont B\kern-0.5em{\scshape i\kern-0.25em b}\kern-0.8em\TeX}}}

\copyrightyear{2023} 
\acmYear{2023} 
\setcopyright{rightsretained} 
\acmConference[CHI '23]{Proceedings of the 2023 CHI Conference on Human Factors in Computing Systems}{April 23--28, 2023}{Hamburg, Germany}
\acmBooktitle{Proceedings of the 2023 CHI Conference on Human Factors in Computing Systems (CHI '23), April 23--28, 2023, Hamburg, Germany}\acmDOI{10.1145/3544548.3581227}
\acmISBN{978-1-4503-9421-5/23/04}

\usepackage{caption,subcaption}
\usepackage{xcolor}

\begin{document}

\ccsdesc[500]{Human-centered computing~Human computer interaction (HCI)}
\title{Fairness Evaluation in Text Classification: Machine Learning Practitioner Perspectives of Individual and Group Fairness}

\rhead{Ashktorab et al.}

\author{Zahra Ashktorab} 
\email{Zahra.Ashktorab1@ibm.com}
\affiliation{%
  \institution{IBM Research}
  \streetaddress{1101 Kitchawan Rd PO Box 218}
  \city{Yorktown Heights}
  \state{NY}
  \postcode{10598}
  \country{USA}
}

\author{Benjamin Hoover} 
\email{benjamin.hoover@ibm.com}
\affiliation{%
  \institution{IBM Research}
  \city{Cambridge}
  \state{MA}
  \postcode{02142}
  \country{USA}
}

\author{Mayank Agarwal} 
\email{mayank.agarwal@ibm.com}
\affiliation{%
  \institution{IBM Research}
  \city{Cambridge}
  \state{MA}
  \postcode{02142}
  \country{USA}
}

\author{Casey Dugan} 
\email{cadugan@us.ibm.com}
\affiliation{%
  \institution{IBM Research}
  \city{Cambridge}
  \state{MA}
  \postcode{02142}
  \country{USA}
}

\author{Werner Geyer} 
\email{werner.geyer@us.ibm.com}
\affiliation{%
  \institution{IBM Research}
  \city{Cambridge}
  \state{MA}
  \postcode{02142}
  \country{USA}
}

\author{Hao Bang Yang} 
\email{hbyang@mit.edu}
\affiliation{%
  \institution{Massachusetts Institute of Technology}
  \city{Cambridge}
  \state{MA}
  \country{USA}
}

\author{Mikhail Yurochkin} 
\email{mikhail.yurochkin@ibm.com}
\affiliation{%
  \institution{IBM Research}
  \city{Cambridge}
  \state{MA}
  \country{USA}
}

\renewcommand{\shorttitle}{Fairness Evaluation in Text Classification.  }

\begin{abstract}
Mitigating algorithmic bias is a critical task in the development and deployment of machine learning models. While several toolkits exist to aid machine learning practitioners in addressing fairness issues, little is known about the strategies practitioners employ to evaluate model fairness and what factors influence their assessment, particularly in the context of text classification. Two common approaches of evaluating the fairness of a model are group fairness and individual fairness. We run a study with Machine Learning practitioners (n=24) to understand the strategies used to evaluate models. Metrics presented to practitioners (group vs. individual fairness) impact which models they consider fair. Participants focused on risks associated with underpredicting / overpredicting and model sensitivity relative to identity token manipulations. We discover fairness assessment strategies involving personal experiences or how users form groups of identity tokens to test model fairness. We provide recommendations for interactive tools for evaluating fairness in text classification.

\end{abstract}

\keywords{human-AI interaction, AI fairness, individual fairness}

\maketitle
\newcommand{\bl}{\textcolor{black}}

\section{Introduction}

 Ensuring an AI system is fair is in the best interest of our society - but beyond that, is oftentimes legally mandated.\footnote{Are you ready for NYC’s anti-bias AI law?} 
 In the past, algorithms have played an active role in furthering social inequities. We have seen that machine learning recognition systems have been less effective in detecting darker skin \cite{krishnapriya2020issues}, search engine results have reinforced representation bias \cite{makhortykh2021detecting}, and translation services misgender women often assigning incorrect male pronouns as the default option \cite{zou2018ai}. These examples show that the lack of fairness in algorithms can negatively impact many aspects of our lives. There has been a flurry of work around algorithmic fairness definitions, algorithms to detect and mitigate bias, and even toolkits with code to easily apply these to new datasets and systems. However, there is a large gap between creating these versus understanding how they are actually reasoned about or applied by the machine learning practitioners in charge of creating production AI systems. 
 
 Algorithmic fairness definitions and methods broadly belong to either group fairness \cite{chouldechova2020snapshot} or individual fairness \cite{dwork2012fairness}. Group fairness requires equitable treatment of groups of people, e.g. comparable loan approval rates for men and women. Regulations based on group fairness are present in banking and are part of the US Equal Employment Opportunity Commission guidelines, known as the 80\% rule \cite{barocas2019Fairness}. One of the most commonly used group fairness definitions is equalized odds \cite{hardt2016equality} which requires class-specific accuracies on minority groups to be similar to the overall performance. These accuracies are easy to measure and can be used to identify group fairness violations. The typical strategy for enforcing group fairness in ML models is to cast it as a constrained optimization problem \cite{agarwal2018reductions}, which is adopted in several toolkits, e.g. Fairlearn and TFCO libraries.

The key idea behind individual fairness is to achieve similar treatment for similar individuals, e.g. candidates that are different only in gender and/or name should receive the same treatment. A prominent example of an individual fairness violation is the 2004 study of the US labor market \cite{bertrand2004emily}, where the investigators responded to job ads with fictitious resumes randomly assigned African-American or white sounding names and observed 50\% more callbacks for the resumes with white sounding names. Identifying individual fairness violations requires comparing outcomes on hand-crafted similar data points as in the aforementioned example, or learning a similarity metric from data \cite{ilvento2019metric,mukherjee2020two} and conducting an individual fairness violation hypothesis test \cite{xue2020auditing,maity2021statistical}. Recent methods for training individually fair ML models cast the problem as an instance of adversarial training \cite{yurochkin2020sensr,yurochkin2021sensei} and are available in the inFairness library \cite{yurochkin2021sensei}. 
 
 Machine learning practitioners are at the forefront of developing  ML models and addressing any potential fairness issues. This is often done iteratively as they develop and test models.
There have been many efforts to help machine learning practitioners more easily address  fairness and bias issues in the form of toolkits that practitioners can use to evaluate the fairness of algorithms. These toolkits allow practitioners to consider fairness mitigation in their modelling work. Such toolkits include IBM’s AI Fairness 360 \cite{bellamy2019ai} and inFairness \cite{yurochkin2021sensei}, Google’s Fairness Indicators \cite{richardson2021towards} and TensorFlow Constrained Optimization (TFCO) \cite{cotter2019two}, Microsoft’s Fairlearn \cite{bird2020fairlearn}, and UChicago’s Aequitas \cite{saleiro2018aequitas}.

There has also been an emergence of tools that aid practitioners in calibrating AI fairness in their models by presenting different types of fairness metrics. Researchers have investigated user perspectives on fairness toolkits \cite{veale2018fairness,richardson2021towards,nakao2022towards}, to understand how to best present fairness metrics to end-users and machine learning practitioners. However, user perspectives of fairness in text classification have been understudied. 

In order to better understand user perspectives, strategies users undertake to determine the fairness of a model, and motivations behind those strategies across different dimensions of fairness such as individual vs group fairness, and how'd they'd prefer to interactively query/work with a system helping them evaluate fairness, we developed a novel interactive tool for investigating fairness metrics for text classification. The tool presents standard accuracy metrics as well as group fairness and individual fairness metrics to users, and allows users to input custom test-cases to interactively determine the fairness of a model. %
We were specifically interested in investigating how the metrics presented to users impact their decisions and the motivations behind their decisions. We were also interested in investigating user strategies in determining the fairness of a model in this interactive environment. 

In this paper, we run an exploratory scenario based study (n=24) in which we present three different models to users and provide different views of the data (overall accuracy metrics, group fairness metrics, individual fairness metrics, as well as opportunities to input custom text to generate metrics for both types of fairness) and ask users to select their preferred model after viewing each metric. We follow with a survey to collect subjective preferences of the metrics they saw, open-ended questions to understand their motivations for the strategies they used during the study, and their perspectives on the metrics with which they were presented. We address the following research questions:

\begin{description}
\item \textbf{RQ1} How do diverse, yet static, metrics impact user perspectives of the models? 
    \begin{description}
    \item \textbf{RQ1a} When viewing standard accuracy metrics? 
    \item \textbf{RQ1b} When viewing group fairness metrics? 
    \item \textbf{RQ1c} When viewing individual fairness metrics? 
    \end{description}
    
\item \textbf{RQ2} What kinds of exploration strategies do machine learning practitioners %
use to determine model fairness when presented with an interactive fairness tool that allows custom inputs? %
 \begin{description}
    \item \textbf{RQ2a} When viewing group fairness metrics? 
    \item \textbf{RQ2b} When viewing individual fairness metrics? 
    \end{description}

\item \textbf{RQ3} How are users' fairness decisions formed as they explore an interactive fairness tool with performance and varying fairness views?
\end{description}

Our contributions in this work are the following:
\begin{itemize}
    \item Introduction of a novel front-end interactive tool that allows exploration of group fairness and individual fairness for text classification.
    \item Identifying strategies users employ to determine the fairness of various models when investigating the tool, including strategies around custom inputs, including the formation of groups for group fairness, and strategies for constructing sentences and identity tokens for individual fairness. 
    \item Based on our findings, we provide design recommendations for future designs of interactive tools to assist machine learning practitioners in reasoning about group and individual fairness. 
\end{itemize}

\section{Related Work}
There has been much research in the CHI, CSCW, and FaccT communities investigating various aspects of the user's relationship with ML fairness and fairness toolkits. From presentation of fairness metrics to understanding organizational challenges within institutions around fairness in AI, there are many opportunities for HCI research.

\subsection{AI Fairness and Decision Making} 

The data used to train machine learning models may conceal discrimination that can be difficult to identify, which has resulted in discrimination and bias in many socio-technical systems and decision-making tasks \cite{berk2021fairness}. There has also been much work on defining algorithmic fairness \cite{calmon2017optimized,friedler2019comparative}. While these definitions can differ in terms of technical details, the intent is similar: equality for protected groups or individuals.
Metrics have been developed to determine whether a decision is fair. For example, demographic parity or statistical parity \cite{yao2017beyond} require minority groups (defined based on, e.g., race or gender) to receive positive outcomes at the same rate as the majority group. \bl{Several prior works discuss the limitations of the algorithmic fairness definitions and metrics, such as the incompatibility of various group fairness notions \cite{kleinberg2016inherent,chouldechova2017fair} and the challenges associated with defining the notion of ``similar individuals'' in the context of individual fairness \cite{fleisher2021s,raz2022gerrymandering}.}

Since AI models and humans work together to make decisions, a human's perceived fairness plays an important role in the decision making tasks \cite{nakao2022towards}. Fairness is highly contextual, and many studies relevant to intelligent user interfaces are interested in how humans make judgments about fairness. 
Prior work has investigated what humans perceive to be fair in various contexts. One study found that non-expert perspectives in the domain of criminal risk and skin cancer prediction judgements about fairness most closely resembled demographic parity \cite{srivastava2019mathematical}. In the context of financial decisions, another study found that end-users ignore individual attributes that may impact the model and ``stereotype'' groups of individuals \cite{woodruff2018qualitative}. Other studies have found that users making judgement favor algorithms that give preferential treatment to protected groups \cite{saxena2019fairness}. In the context of a study on recidivism, disclosing the race of defendants had an impact on users decisions about defendants \cite{mallari2020look}.

\textbf{Fairness in Toxic Text Classification.} Bias in toxic text classification often occurs when the identity tokens in a training dataset are disproportionately used in toxic comments. An example of this is if the word ``gay’’ appears in toxic comments a lot more often than it appears in other comments, resulting in a model that makes generalizations about different words, such as tying the word ``gay’’ to toxicity. Dixon et al. manually identify a set of 50 common identity tokens (including but not limited to: atheist, queer, feminist, black, muslim) that disproportionately appear in toxic comments in training datasets \cite{dixon2018measuring}. 
In this work, we utilize the same toxic comment classification dataset and build our fairness metrics using their set of 50 identity tokens.
To alleviate bias in toxic text classification, researchers have considered various group fairness methods \cite{soares2022your,ball2021differential,Pruksachatkun2021} where the goal is to enforce comparable accuracy across comments referring to identities or containing identity tokens to achieve equalized odds \cite{hardt2016equality}. Garg et al. pose the question, ``How would the prediction change if the sensitive attribute referenced in the example were different?'' \cite{garg2019counterfactual} and propose a corresponding metric based on a comparison of the predictions on sentences that only differ in the identity token. A similar metric was used in prior work on achieving individual fairness in toxic text detection \cite{yurochkin2021sensei,petersen2021post}. The fairness metrics in our tool are based on these works.
\bl{Beyond toxic classification, there are concerns about the disparate harms of NLP technologies on different demographic groups \cite{prost2019debiasing}. Researchers have explored bias, or ``a skew that produces a type of harm’’ \cite{crawford2017trouble} in the context of different NLP tasks. Biased metrics or group fairness measures are used to show the differences between demographic groups produced by a model. These NLP tasks include question answering \cite{li2020unqovering}, relation extraction \cite{gaut2019towards}, text classification \cite{de2019bias}, autocomplete generation \cite{sheng2019woman} and machine translation \cite{stanovsky2019evaluating}. In these tasks, demographic dimensions are an identity axis on which they are evaluated for bias. Many of these tasks result in bias along various demographic dimensions including: gender/nationality/age/religion/race through identity terms \cite{dev2022measures}. }

\subsection{ML Practitioner Perspectives of Fairness} 
Prior research has highlighted the gap between institutional and ML practitioners' goals regarding fairness, revealing a significant disconnect. Institutional initiatives to ensure fairness are frequently led by individuals passionate about transparency and equity. However, reconciling the drive to meet product goals while remaining committed to fairness and transparency can present challenges \cite{john2022reality}. Additionally, research examining AI fairness in high stakes decision-making has revealed that practitioners have difficulty identifying the most relevant direct stakeholders and demographic groups on which to focus when considering fairness metrics \cite{10.1145/3512899}. Practitioners also struggle to explain that performance metrics like accuracy are not always appropriate for evaluating fairness, especially when dealing with imbalanced data \cite{veale2018fairness}.

Researchers have explored the impact of different design strategies on users' insights when using bias detection prototypes. One study found that the comprehensiveness and information load of toolkits are crucial considerations for optimal design \cite{richardson2021towards}. In addition, research has highlighted the importance of fairness-aware data collection, the challenge of identifying subgroups that are most adversely affected by model bias, and the misalignment of tools and practices with practitioners' workflows and organizational incentives \cite{holstein2019improving, madaio2020co}. While toolkits can guide practitioners towards fairer models, they require additional functionality and resources. Studies have developed rubrics with criteria for optimal tooling, such as providing global and local perspectives of fairness, well-supported demos and tutorials, and an explicit interpretation of tool limitations \cite{madaio2020co}. Other research suggests that fairness toolkits should broaden their scope, be designed for interdisciplinary research, and include anti-patterns in addition to patterns for using toolkits \cite{10.1145/3531146.3533113}. Furthermore, practitioners may compare individual instances in addition to group metrics when evaluating model fairness, as found in a recent study on loan application evaluations \cite{nakao2022responsible}.

Most recently, researchers investigated how ordinary end-users can explore interpretable interfaces through an Explainable AI approach called "explanatory debugging" to identify potential fairness issues. They created a prototype of such a system and evaluated how cultural dimensions developed by Hofsted et al. (i.e., power distance, individualism, masculinity, uncertainty avoidance, long-term orientation, and indulgence) \cite{hofstede2011dimensionalizing} impact the use of this tool \cite{nakao2022towards}. Prior work has identified challenges associated with considering fairness metrics, identifying the most relevant stakeholders, and aligning tools with practitioners' workflows and organizational incentives. Our study aims to expand upon this prior research by exploring user perspectives of model fairness when presented with interactive fairness tools that allow custom inputs. We seek to understand the impact of diverse metrics on user perspectives of models and the strategies used by machine learning practitioners to determine model fairness when presented with such tools. Additionally, we aim to understand how users' fairness decisions are formed as they explore interactive fairness tools with varying fairness views.

\section{Methodology} 
To understand how practitioners interacted with fairness metrics, we conducted an exploratory experiment followed by a survey. Our experience consisted of views of 1) performance metrics (Figure \ref{fig:images}), 2) group fairness metrics (Figure \ref{fig:group}) and 3) individual fairness metrics (Figure \ref{fig:individual}) followed by an exploration view (described in Section \ref{section:explore}), in which users could provide custom inputs to further inspect the models. Users were told they were designing and implementing a new social media platform for posting announcements through free form text for colleagues and friends. We instructed them, ``In the past, there have been issues with people posting toxic comments. Toxic comments are comments intended to demean a person or groups of people. Your colleagues have considered this problem and have implemented three toxicity classifiers to catch and flag any text that might be toxic. You must decide which model to deploy.'' 

After being given instructions, users went through three different pages in which they were provided with additional metrics on the models (Model A, Model B, and Model C) and asked to choose which model they would deploy after seeing each page.

\subsection{Task 1: Overall Performance} 
\begin{figure*}
    \centering %
    \begin{subfigure}{0.64\textwidth}
  \includegraphics[width=\linewidth]{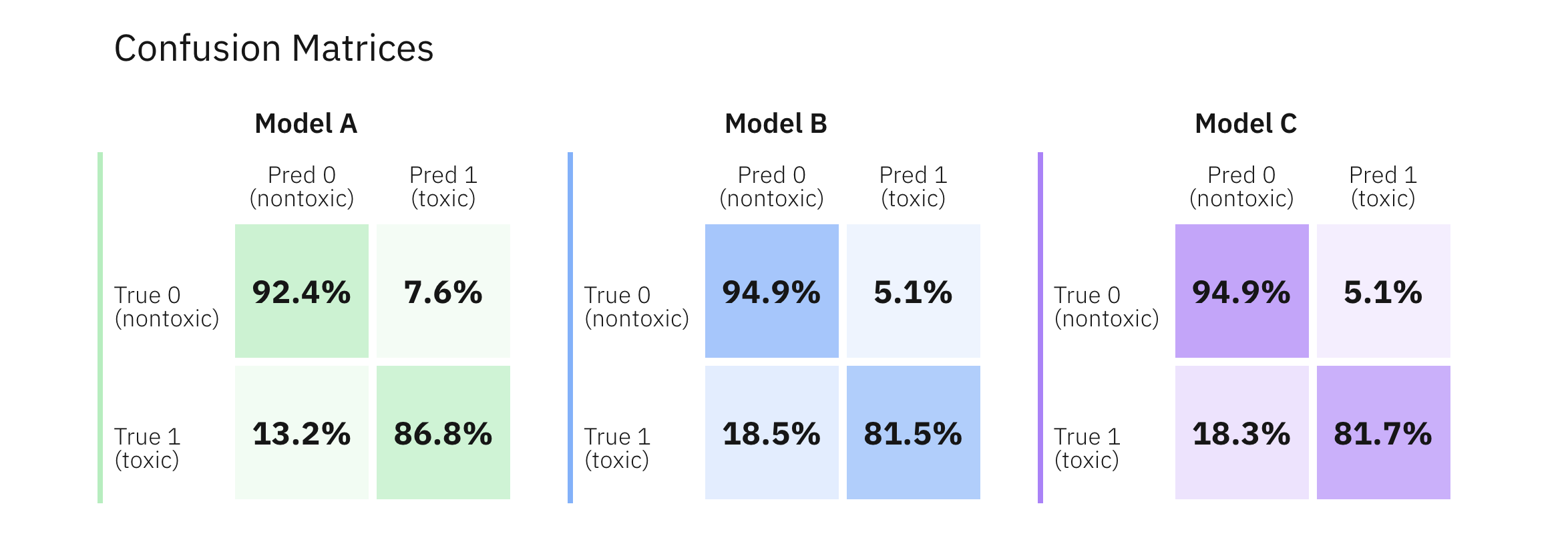}
  \caption{Confusion Matrices for each of the models showing true positives, true negatives, false positives, and false negatives.}
  \label{fig:3}
\end{subfigure}\hfil
\begin{subfigure}{0.34\textwidth}
  \includegraphics[width=\linewidth]{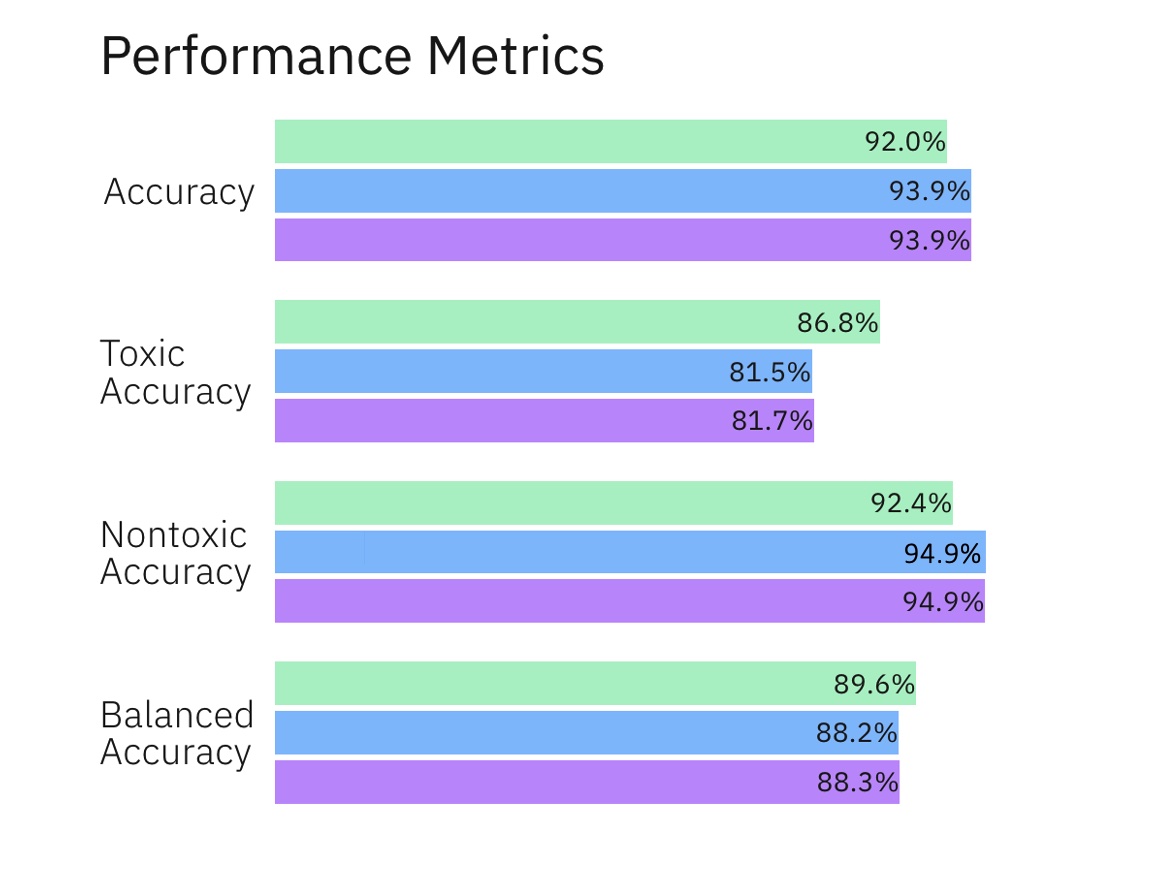}
  \caption{Accuracy, Toxic Accuracy, Non-toxic accuracy, and Balanced Accuracy metrics for each of the models.}
  \label{fig:1bar}
\end{subfigure}\hfil %

\caption{Accuracy metrics shown to users in Task 1. }
\label{fig:images}
\end{figure*}

Our study began by showing users accuracy metrics of the entire model, for each of the three models. Our aim was to explore how users determined the best model to deploy without being presented with the fairness metrics. We provided the following prompt to guide users about their task: 

\begin{quote}
\textit{You begin by examining various performance metrics of the three models. You notice that the majority of the comments (about 90\%) are non-toxic, thus a simple accuracy metric may not be sufficient. For example, a classifier always predicting "non-toxic" will have an impressive 90\% accuracy, yet it is useless. Therefore you also consider accuracy on only toxic comments, accuracy on only non-toxic comments, and balanced accuracy (average of the former two metrics).}
\end{quote}
We provided users with various accuracy metrics (accuracy, toxic accuracy, non-toxic accuracy, and balanced accuracy) as well as confusion matrices for each of the models as seen in Figure \ref{fig:images}. A good toxicity classifier should have good performance in terms of all of these metrics. A classifier always predicting ``non-toxic'' will have 90\% accuracy and 100\% non-toxic accuracy, but 0\% toxic accuracy, and only 50\% balanced accuracy. After seeing these metrics, users selected the best model to deploy.

\subsection{Task 2: Performance on Group Filtered Views of the Data}
\begin{figure*}[htbp]
\centering
\includegraphics[width=0.8\textwidth]{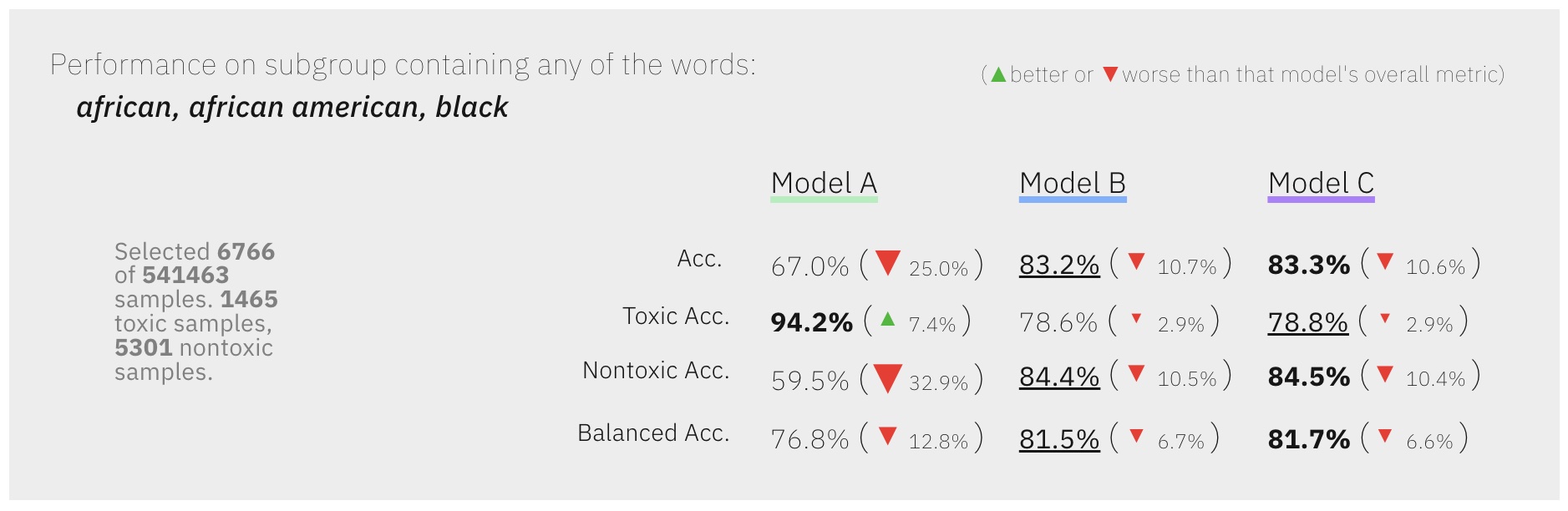}
\caption{Group View: For a selected subgroup, each model's performance difference, compared to its performance on the whole dataset, is presented. (For each example, the model that performs best on a given accuracy metric is written in bold, and the second best is underlined). Red (green) triangles, reminiscent of the markers beside familiar stock tickers, indicate the decrease (or improvement) of the model's performance on that subgroup, compared to the overall dataset, where marker size is scaled by the difference in metrics. Note that to satisfy group fairness, a performance of a model on a subgroup should be similar to its overall performance (i.e., the size of the markers for all metrics should be small).}
\label{fig:group}
\end{figure*}

In Task 2, we presented users with a task to check for algorithmic bias. Users were shown subgroups of data based on words describing minority identities. The subgroups are defined by filtering the dataset for examples containing particular words. We selected groups of words related to various identities such as sexuality, race, ethnicity, and religion. An example of one such subgroup can be seen in Figure \ref{fig:group} in which the text has been filtered for mentions of ``african'', ``african american'', and ``black''. Other groups we presented to users are ``lesbian,gay, bisexual, queer, lgbt, lgbtq, homosexual'' and
``blind, deaf, paralyzed.'' \bl{We use group identifier keywords as a proxy for text content, which is a common way to find groups of text in text classification \cite{dixon2018measuring}. We later describe the limitations of this approach and alternative ways to find groups within text.} We present users with the same metrics as seen in Task 1 (accuracy, toxic accuracy, balanced accuracy) using the original bar charts seen in Figure \ref{fig:1bar} as well as icons next to the metrics for each group representing how they compare relative to the model's overall metrics (Figure \ref{fig:group}). After seeing this view, users selected the model they would deploy.

\subsection{Task 3: Performance on Individual Fairness Sentences in the Data}
\begin{figure*}[htbp]
\centering
\includegraphics[width=0.8\textwidth]{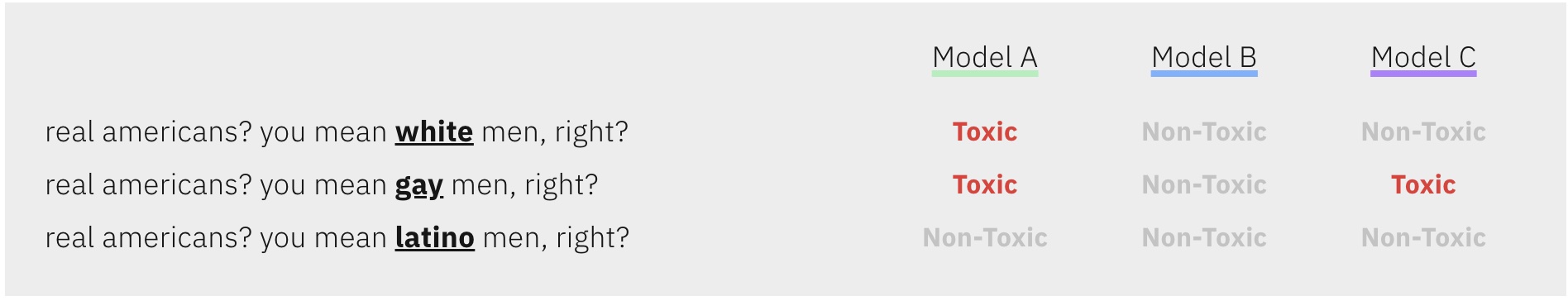}
\caption{Individual View: Given one sentence, a user is able to see the classifications (Toxic, Non-Toxic) for each model (Model A, Model B, Model C) when the identity token is replaced.}
\label{fig:individual}
\end{figure*}

In Task 3, we allowed users to examine the following four sentences, and see how the model changed its classification when identity tokens appearing in brackets below in the sentence were changed. 
\begin{itemize}

\item congratulations to 2 fine [young, gay, black] men! go beavers! 
\item real americans? you mean [white, gay, latino] men, right?
\item name one city in the western world where [deaf, black, transgender] people are truly respected
\item a sad day for [lesbian, white, american] people everywhere
\end{itemize}
Users observed the impact of manipulating identity tokens on the classifications made by Model A, Model B, and Model C. We asked participants to review examples and select the model to deploy after seeing several sentences (listed above) with various identity tokens and the respective model classifications for each model. An example can be seen in Figure \ref{fig:individual}.

\subsection{\bl{Task 4:} Exploration View} 
\label{section:explore}
\begin{figure}
  \includegraphics[width=\linewidth]{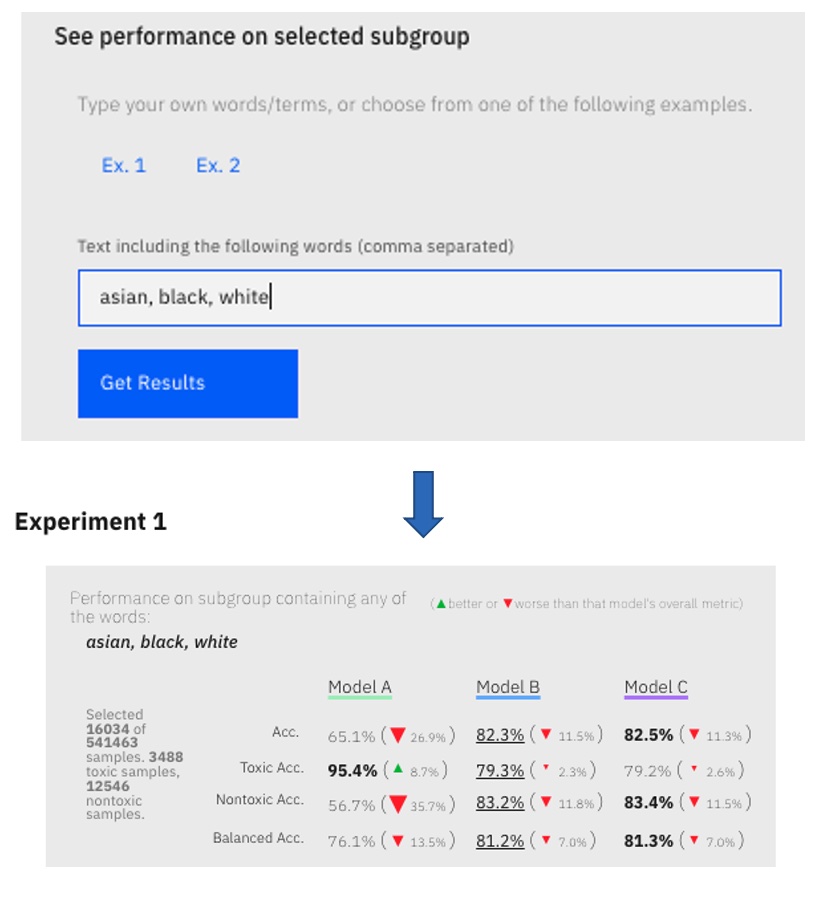}
  \caption{Interface for custom inputs for Group fairness with resulting output. }
  \label{fig:customg}
\end{figure}

\begin{figure}
  \includegraphics[width=\linewidth]{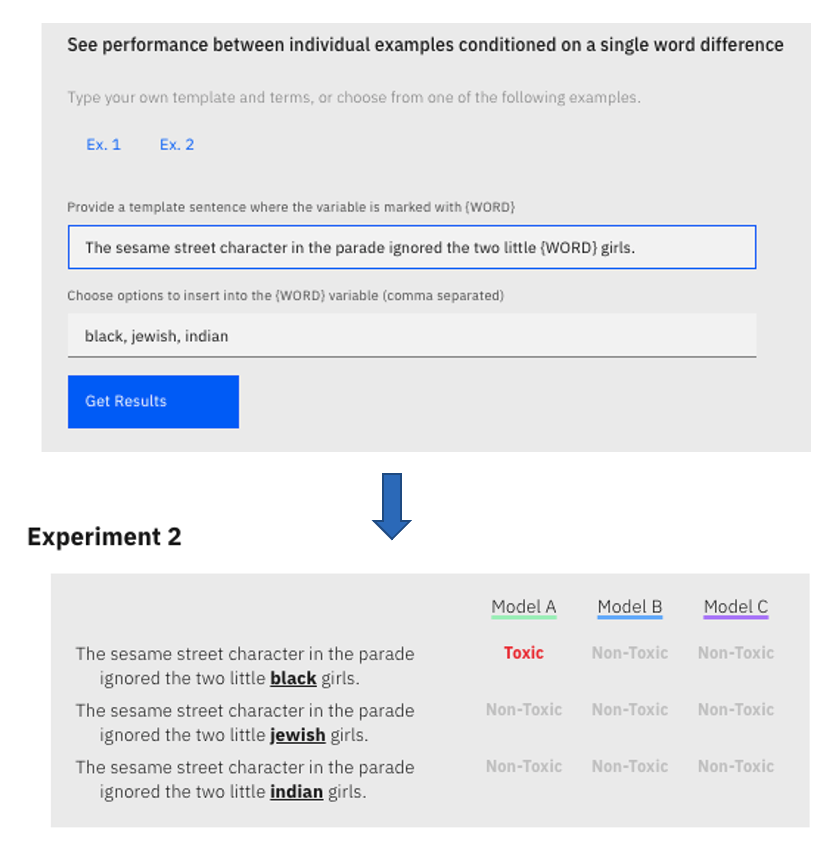}
  \caption{Interface for custom inputs for Individual fairness with resulting output.}
\label{fig:customi}
\end{figure}

In order to further answer our research questions, we designed and developed a tool that would allow users to interactively test group fairness metrics and individual fairness metrics of models for toxic text detection. Users were able to experiment with four different types of interactions (1) Pre-populated Group View, (2) Pre-populated Individual View, (3) Custom Group View, and (4) Custom Individual view (See Figures \ref{fig:customg} and \ref{fig:customi}). Before introducing users to the interactive portion of the study, we first had them complete three tasks that acquainted them with the accuracy metrics, group fairness, and individual fairness metrics. In the exploration phase, participants had the chance to inspect the individual and group views presented to them in the previous tasks, but with their own custom inputs. To guide them on the usage of our tool, we pre-populated the exploration views with some examples. 

\begin{figure*}[htbp]
\centering
\includegraphics[width=0.7\textwidth]{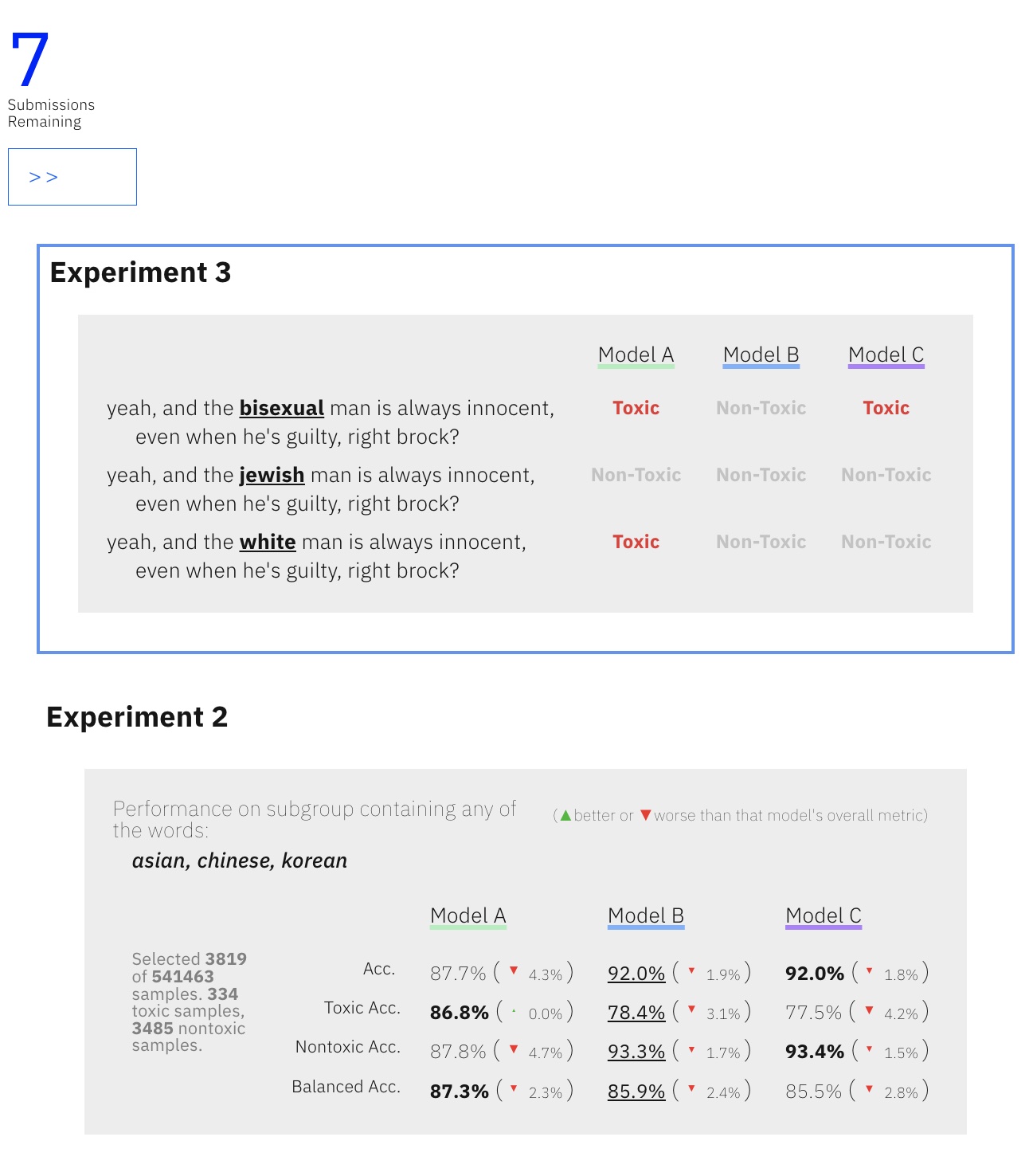}
\caption{Users are asked to run 10 ``experiments'' in the Exploration View. They can either test different subgroups by selecting words, or test example sentences by typing their own template. This image shows an example of two experiments done, one for the group view and another for the individual view. This user has 7 experiments remaining.}
\label{fig:exploration}
\end{figure*}

Participants were instructed to conduct 10 additional model checks to ultimately make a decision as to which model to select. They could choose to either see the performance of a selected subgroup or see the performance between examples conditioned on a single word difference. An example of this view can be seen in Figure \ref{fig:exploration}. Participants were able to do the 10 experiments (group-view or individual-view) either with our pre-populated examples or with examples of their own. At the conclusion of the experiments, users were given the chance to select a final model. Figures of the interface that prompted users for custom examples can be seen in Figure \ref{fig:customg} and \bl{Figure \ref{fig:customi}}. The fairness results appeared on an Experiment panel on the right hand-side of the page, that could be minimized or maximized depending on user preference. At any point, users were able to scroll through the panel to see a view of all of the experiments that they had run on the Exploration page. 
Below, we describe each of the views with which users interacted. 
\begin{description}
\item \textbf{Pre-populated Group View} The pre-populated group view presented metrics (accuracy, toxic accuracy, nontoxic accuracy, and balanced accuracy) for the pre-defined groups we provided for users. Similarly to the Group View in Task 2, we showed users how many total documents were selected from the dataset based on the defined groups, the number of toxic samples, and the number of nontoxic samples. There were also indicators next to each performance metric to identify if there was an increase/decrease in the corresponding metric as seen in Figure \ref{fig:group}.  
\item \textbf{Pre-populated Individual View} The pre-populated individual view presented the classification for one sentence in which one word was replaced with different identity tokens, similarly to the Individual View in Task 3. For each model and sentence variation, users were presented with either a ``Toxic'' classification or a ``Nontoxic'' classification as seen in Figure \ref{fig:individual}. 
\item \textbf{Custom Group View} The custom group view was similar to the pre-populated group view, except users were able to define their own groups across the validation set. Identity tokens were separated by commas. An example of an experiment can be seen in Figure \ref{fig:customg}.
\item \textbf{Custom Individual View} The custom individual view was similar to the pre-populated individual view except users could formulate their own sentences and choose one word which would be replaced with other identity-token words of their choosing. Users could write a sentence including the term ``\{Word\}'' to indicate where identity tokens of their choosing would be substituted in (indicated as a separate list, separated by commas). An example of this input interaction can be seen in Figure \ref{fig:customi}.
\end{description}

\subsection{Model Descriptions} 
\label{sec:modeldescriptions}

In our experiments, we presented metrics for three different models that behaved differently. \bl{We trained all models using the ``Unintended Bias in Toxicity Classification'' dataset \cite{dixon2018measuring}.\footnote{\url{https://www.kaggle.com/c/jigsaw-unintended-bias-in-toxicity-classification}} See Appendix \ref{apdx:dataset-statistics} for dataset details.} Below, we describe the models and why they behaved differently. 
\subsubsection{Model A}
Model A is a ``standard'' model. We utilized the standard practice in today's natural language processing and fine-tuned a BERT (base, uncased) model \citep{devlin2018bert} using the HuggingFace library \cite{wolf2019huggingface}. As discussed previously, due to class imbalance in the data (the majority of the comments are non-toxic), we prioritized balanced accuracy by enforcing an equal number of toxic and non-toxic samples in the mini-batches. This aligns with the previous studies of this dataset \cite{yurochkin2021sensei}.

\subsubsection{Model B}
Model B is an ``individually fair'' model. Its goal is to achieve similar treatment of similar inputs, where similar inputs are sentences with similar context, but potentially different identity mentions, e.g. ``My friend is American'' and ``My friend is gay''. In total, we took 50 identity tokens into consideration that were previously recommended for measuring bias in toxic text classification \cite{dixon2018measuring}. We \bl{used} fine-tuned BERT to achieve both balanced accuracy and individual fairness using the SenSeI algorithm from the inFairness package \cite{yurochkin2021sensei}, similar to the original study of individual fairness in toxicity classification \cite{yurochkin2021sensei}.

\subsubsection{Model C}
\label{sec:models:group}
Model C is a ``group fair'' model. The goal of group fairness is to achieve equal treatment of groups of inputs. We trained this model following the corresponding example provided in the TFCO library.\footnote{\url{https://github.com/google-research/tensorflow_constrained_optimization/blob/master/examples/colab/Wiki_toxicity_fairness.ipynb}} The groups are defined based on the presence of identity tokens. For example, one group consisted of all sentences where at least one of the words ``lesbian'', ``gay'', ``bisexual'', ``queer'', ``lgbt'', ``lgbtq'', ``homosexual'', ``straight'', ``heterosexual'' was present and another group of all sentences with any of ``christian'', ``muslim'', ``jewish'', ``buddhist'', ``catholic'', ``protestant'', ``sikh'', ``taoist''. There were a total of six groups. The equitable treatment of groups was ensured by constraining the balanced accuracy of each group to be $\geq 95\%$ (on train data). For this model, we fine-tuned BERT with the aforementioned group fairness constraints using TFCO library (note that the constraints also account for class imbalance).

We summarize the performance and fairness metrics corresponding to the three models in Table \ref{tab:model-metrics}. Model A, as expected, has the highest accuracy and the worst fairness metrics. Model B and Model C are slightly less accurate but are noticeably fairer. In this case, training for either individual or group fairness resulted in the improvement of both types of fairness metrics. However, as we shall see in our analysis, the model selected most commonly in each of the views corresponded to the model trained for the metric emphasized by the view (accuracy, group, and individual).

\begin{table*}[]
\centering
\begin{tabular}{|c|c|c|c|}
\hline
Model & Balanced Accuracy (BA) & Prediction Consistency & BA difference on minority groups \\ \hline
Model A & 89.61\% & 49.70\% & 6.04\% \\ \hline
Model B & 88.29\% & 73.62\% & 4.51\% \\ \hline
Model C & 88.3\% & 72.25\% & 4.60\% \\ \hline
\end{tabular}
\caption{Performance and fairness metrics for the three modes presented in our study. Balanced accuracy (BA) quantifies classification performance accounting for class imbalance; Prediction consistency is the individual fairness metric and is the frequency of comments where prediction does not change when varying the identity token; BA difference on minority groups is the group fairness metric (smaller is better) and is the average absolute difference between BA on minority groups and the overall BA.}
\label{tab:model-metrics}
\end{table*}

\bl{We provide more details about the training procedure of these models along with the hyperparameter settings in Appendix \ref{apdx:model-training}.}

\subsection{Participants}
We recruited participants across four US-based institutions (Academia and Industry) in July and August of 2022. Recruiting happened via email, word of mouth, and snowball sampling. 
We asked participants about their prior exposure to AI. 17 of the participant\bl{s} reported having ``some work experience and/or formal education related to AI'', \bl{7} of the participants reported having ``extensive experience in AI research and/or development'', \bl{and 4 of the participants reported ``no prior exposure to AI''.} Participants who reported not having exposure to AI \bl{and those did not complete the entire study} were not included \bl{ in our analysis which left us with 24 participants}. We also asked participants about their experience with AI fairness. \bl{12} participants reported having some work experience and/or formal education related to AI fairness. Four users reported that they closely follow AI-fairness related news, 6 users reported that they had heard about AI fairness from the news, friends and/or family, and \bl{two} user\bl{s} reported having extensive experience in AI fairness research. 

Participants were asked about their area of work and were able to select multiple areas because of the overlap of ML technologies. The areas spanned NLP (8), Predictive analytics (7), Conversational AI/Chatbots (7), Decision Support Systems (4), Information Retrieval (3), Computer Vision (3),  Human computer interaction/Human Centered AI (3), User Modeling (2), Speech and Voice (1), Robotics (1),  Recommendation Systems (1), and Statistics (1).

\section{Results} 
\begin{figure}[htbp]
\centering
\includegraphics[width=\linewidth]{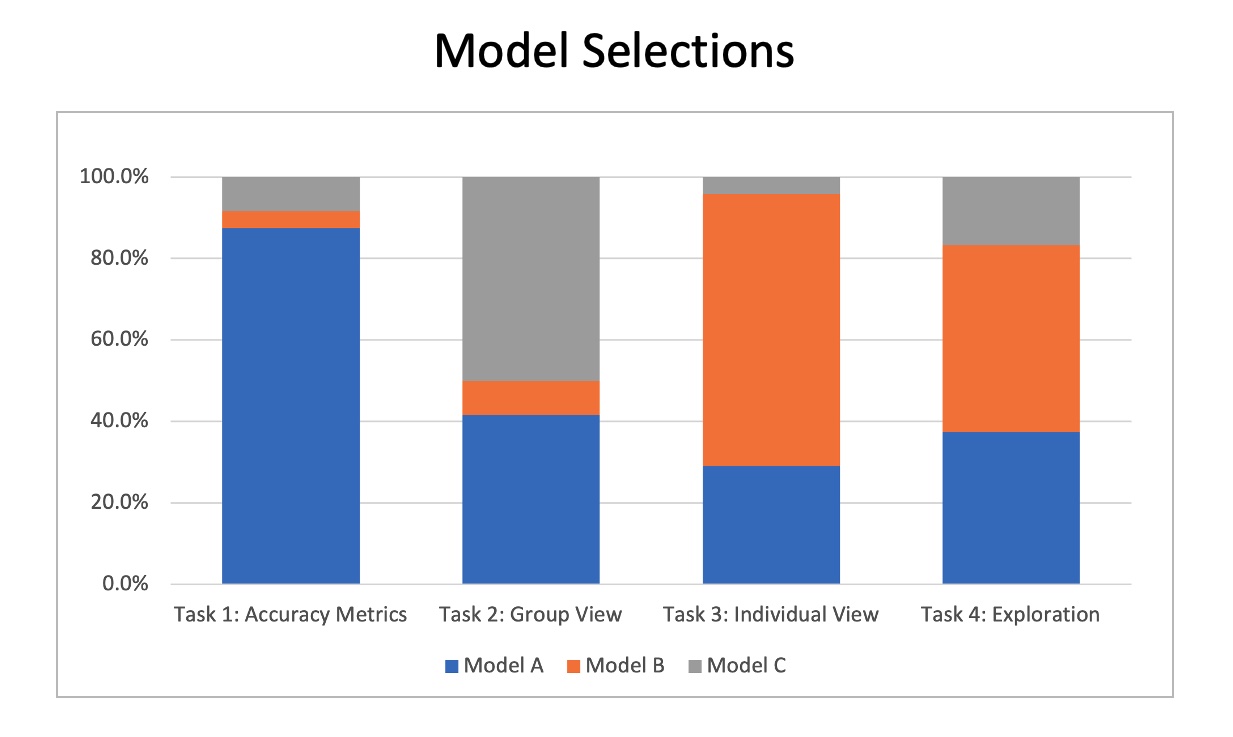}
\caption{\bl{Models selected after each phase by the 24 participants. After Task 1, majority of participants chose Model A. After Task 2, participants were split between Model A and Model C. After Task 3, Majority of Participants chose Model B. After exploring all of the models, the participants were split between Model A and Model B.}}
\label{fig:selec}
\end{figure}

We report on what model was selected after each task in Figure \ref{fig:selec}.
We also examined the open-ended feedback and for each model selection, we identified why the majority of users selected that model. \bl{For all of the themes, two of the authors independently reviewed the reasons to extract themes for model selection for all four of the tasks. After independently generating codes, the authors consulted one another and iterated over codes until they agreed on all of themes generated.} We describe the emerging themes in the next section. In Section 5.4 we then report our findings on how our subjects used the exploration views as a tool to assist in making model selections.

\bl{Before discussing the results for each task, we provide definitions of the emerging themes listed in Table \ref{tab:reasons} below:}
\begin{itemize}
 \item \bl{\textbf{Prioritizing Toxic Accuracy}: Participants prioritized maximizing percentage of instances correctly classified as toxic.}
 \item \bl{\textbf{Prioritizing Non-Toxic Accuracy}: Participants prioritized maximizing the percentage of instances correctly classified as non-toxic. }
 \item \bl{\textbf{Prioritizing Balanced Accuracy}: Participants prioritized maximizing the average percentage of instances correctly classified as toxic and the percentage of instances correctly classified as non-toxic. }
 \item \bl{\textbf{Toxic Accuracy Detection for Minority Groups}: Participants prioritized maximizing percentage of instances correctly classified as toxic for either pre-defined minority groups or minority groups they explored in custom phase.}
 \item \bl{\textbf{Best Overall Metrics}: Perception that all four of the combined metrics  provided for group view (accuracy, toxic accuracy, non-toxic accuracy, and balanced accuracy) is ``better'' for one model over another.}
 \item \bl{\textbf{Identity Changes Impact Model Classification}: Replacing an identity token in a sentence in individual fairness view changed the classification (toxic vs. non toxic) of the model.}
\item \bl{\textbf{Identity Changes do not Impact Model Classification}: Replacing an identity token in a sentence in individual fairness view did not change the classification (toxic vs. non toxic) of the model (i.e. ``I went to the mall with my gay friend'' and ``I went to the mall with my straight friend'' yielded the same classification.)}
\end{itemize}

\begin{table*}
\small
\centering
\begin{tabular}{p{1.5cm}|p{2cm}|p{4cm}| p{7.5cm}} 
View &Model Selection&Reason \bl{(Percentage of Those Who cited this Reason)} &Example Quote\\
\hline
\hline
Overall Accuracy&\textbf{Model A (88\%)}& Prioritizing Toxic Accuracy \bl{(81\%)}& Prioritizing recall here, we would rather overpredict toxicity than underpredict.\\\cline{3-4}
&&\bl{Prioritizing Balanced Accuracy (19\%)} & \bl{Because model A has a more balanced accuracy for toxic and nontoxic classes. So, I believe that model A is more fair than the others.} \\\cline{2-4}
&Model B (4\%)&Prioritizing Non-Toxic \bl{ Accuracy (100\%)}& Model B has the highest score for Non-Toxic accuracy. I think it is better to miss some Toxic posts than to incorrectly flag a Nontoxic post as Toxic. \\\cline{2-4}
&Model C (8\%) & \bl{Prioritizing Non-Toxic Accuracy(100\%)}&Model C is definitely better than B, because it has the same false positive rate and a lower false negative rate. There is a compromise between A and C, where A predicts a higher proportion of true toxic comments but also has a higher false positive rate. I chose model C because it seems to most accurately reflect content, leading to a more accurate overall picture, and also has a lower false negative rate, which negatively impacts 'good' users.\\\cline{4-4}
&&&\bl{Many platforms today have histories of marking "toxic" content when it actually isn't, and that's led to some pretty discouraged or angry users who find it frustrating that their non-toxic content was taken down. No model is going to be perfect, but I think it's better to have a model that has higher rates of detecting non-toxic material to best prevent false positives from happening.}\\\hline
Group View &Model A (42\%)& Toxic Accuracy Detection for Minority Groups \bl{(100\%)}& I think I value a more a fair model compared to others. We have a great recall (detecting toxic comments) increase for minorities hurting the general accuracy by only 2.5 points (task 1)\\\cline{2-4}
&Model B (8\%) & Best Overall Metrics \bl{(100\%)}& If we look at the subgroups metrics, model B has the best metrics for 2 groups generally. So, that is the reason to pick model B.  \\\cline{2-4}
&\textbf{Model C (50\%)}& Prioritizing Non-Toxic Accuracy \bl{(58\%)}&While Model A performs best on Toxic comments, it is also clearly flagging many normal comments that mention a minority group as toxic.  Model C seems to help alleviate this problem best.\\\cline{3-4}
&&\bl{Prioritizing Balanced Accuracy (42\%)} & \bl{The balanced accuracy is relatively high in model C across all subgroups.} \\\hline

Individual View&Model A (29\%)& Identity Changes Impact Model \bl{classification (100\%)} &Model B predicts everything as nontoxic. Also, I think most of those sentences are a little toxic. So, I believe model A is better than the others.\\\cline{2-4}
&\textbf{Model B (67\%)}& Identity Changes Do not Impact Model \bl{classification (100\%)}& Model B is most resistant to the effect of identity changes in a sentence. \\\cline{2-4}
&Model C (4\%)& \bl{Prioritizing Balanced Accuracy (100\%)} &Model C seems to be the most balanced despite its errors. Model A seems to be producing too many false positives on toxicity, which could undermine the cause of fairness more than support it.\\\hline
Exploration View&Model A (38\%) &\bl{Prioritizing} Toxic Accuracy \bl{(76\%)} &Actually, all of these models are not handling fairness issue for Asians, Africans, etc. I choose the model having highest toxic accuracy.\\\cline{3-4}
&&\bl{Prioritizing Balanced Accuracy (14\%)}&\bl{I found balanced accuracy the easiest to think through}\\\cline{2-4}
&\textbf{Model B (46\%)}&Identity Changes Do not Impact Model \bl{classification (100\%)}&Best balanced accuracy with minority subgroups and more consistent classifications despite different identity tokens\\\cline{2-4}
&Model C (16\%)&\bl{Prioritizing} Balanced \bl{Accuracy (75\%)} &Writing more nuanced sentences on controversial topics, I affirmed my choice of Model C based on the fact that it seems more balanced in identifying toxicity. I value free speech as well as safe space, so I seek a model that is balanced. \\\cline{3-4}
&&\bl{Toxic Accuracy for Minority Groups (25\%)}&\bl{Model A does not classify non-toxic comments that mention minority or disadvantaged groups well.  Model C does better at this while still giving a fair performance on toxic comments that mention such groups.}\\
\hline
\end{tabular}

\caption{\label{tab:reasons}Emergent themes as to why people selected a model after each phase of the experiment. For each view, the most selected Model is bolded.}
\end{table*}%

\subsection{Task 1: Accuracy}
Our participants were first presented with accuracy metrics of the models and then asked to choose the model they would deploy. Majority of respondents (88\%) selected Model A after just viewing accuracy metrics. \bl{Table \ref{tab:reasons} lists the reasons users cited why a particular model was selected with the corresponding number of participants who cited that reason. } \bl{Two reasons emerged for selection of Model A over the others cited in Table \ref{tab:reasons}: Prioritizing Toxic Accuracy and Prioritizing Balanced Accuracy. The participants who selected Model B cited that they did so to prioritize non-toxic accuracy, and all participants who selected Model C did so because they felt the model most accurately reflects real world content and classifications.} For example, while one participant who selected Model A said, ``We would rather overpredict toxicity than underpredict'', another participant who selected Model B said ``I think it is better to miss some Toxic posts than to incorrectly flag a Nontoxic post as Toxic.'' The participants who prioritized toxic accuracy (the majority) selected Model A, while \bl{the three participants who cited prioritizing} non-toxic accuracy \bl{selected} Model C (8\%), then Model B (4\%). \bl{Model C and Model B both have the same non-toxic accuracy (94.9\%).}
When viewing the standard metrics, \bl{most} users focused on toxic accuracy, preferring to overpredict toxicity to capture all instances of toxicity rather than underpredicting toxicity (\textbf{RQ1a}).

\begin{quote}
    \textit{Model B has the highest score for Non-Toxic accuracy. I think it is better to miss some Toxic posts than to incorrectly flag a Nontoxic post as Toxic.}  Participant \#33 (Task 1 selection: Model B, Task 2 selection: Model C, Task 3 selection: Model B, Exploration selection: Model B)
\end{quote}

\subsection{Task 2: Group View}
 For group view, 12 participants \bl{(50\%)} selected Model C, 10 \bl{(42\%)} participants selected Model A, and 2 participants \bl{(8\%)} selected Model B. \bl{All of the participants who selected Model A cited toxic accuracy as the reason for their selection. Both participants who selected Model B cited that Model B had the best overall metrics. Of the individuals who selected Model C 58\% cited that they were prioritizing non-toxic accuracy and 42\% cited that they were considering Balanced accuracy when making their selections.} 
Overall, the preferences after seeing group view were split between Model C (50\%) and Model A (42\%), with only 2 participants selecting Model B.  

As described in Section \ref{sec:modeldescriptions} on Model Descriptions, Model C was specifically optimized for group fairness. It appears that when viewing group fairness metrics, participant responses slightly favored the group fair model (Model C) (50\%), but were also split with selecting the standard model (Model A) (42\%). Those who selected Model A, again focused on toxic accuracy for the minority groups, noting the increase in toxic accuracy for the presented groups. Those who selected Model C, prioritized non-toxic accuracy \bl{and balanced accuracy} taking into account normal comments that mention minority groups that may be flagged as toxic (\textbf{RQ1b}).

\begin{quote}
\textit{Model A has good toxic accuracy, but bad nontoxic accuracy, leading to non toxic messages being removed unnecessarily. Therefore model C performs better on subset 1 + 2. Model B performs better on subset 3, but as this is a smaller subset, I'd choose model C. } \small{Participant \#36 (Task 1 selection: Model A, Task 2 selection: Model C, Task 3 selection: Model B, Exploration selection: Model C)}
\end{quote}

\subsection{\bl{Task 3:} Individual View}
After seeing the individual view, the majority of the participants selected Model B as the model they would choose to deploy (67\%) \bl{because it was resistant to change with identity tokens}, followed by Model A (29\%), followed by Model C (4\%). Those who selected Model B believed that replacing words that represent a person's identity in the sentences should not change the toxic classification of the sentence. Users who selected Model B said they did not believe any of the sentences we presented to them were toxic regardless of the identifiers which were replaced in the sentences. 
\begin{quote}
    \textit{I don't think any of the above comments are toxic, so Model B has the highest accuracy in my opinion.}  Participant \#14 (Task 1 selection: Model A, Task 2 selection: Model A, Task 3 selection: Model B, Exploration selection: Model B)
\end{quote}

 When observing the individual view, the majority of the participants focused on the impact of identity changes to the model. Those who selected Model B (the individually fair model) did so because the identity token changes in the sentences did not impact the outcome of the model. \bl{The participants who selected} Model A cited the impact of the identity token changes in sentences but believed identity token sentences should impact a model's classification output (\textbf{RQ1c}).

\begin{quote}
\textit{Model A may seem like overkill but it actually seems to be the one catching actually toxic statements.} Participant \#38 (Task 1 selection: Model A, Task 2 selection: Model A, Task 3 selection: Model A, Exploration selection: Model A)
\end{quote}

\subsection{\bl{Task 4:} Exploration View} 

\bl{In the Exploration view, }Model B was the most preferred model with 46\% of participants choosing it. 38\% of participants selected Model A, and only 4 participants (17\%) selected Model C. For Model B,  \bl{the} participants noted that they believed that identity tokens should not change classifications and felt that Model B was most resistant to change. Those who selected Model A believed that the other models were not handling fairness issues for minorities \bl{and prioritized toxic accuracy and balanced accuracy}. \bl{Participants selected Model C because they prioritized balanced accuracy and felt that it performed best for minority groups. All of the themes can be seen in Table \ref{tab:reasons}.}
Overall, there were 4 decisions on model deployment users needed to make, and which model was selected by the majority differed at each decision point, showing that a given view can highly influence decision-makers. At the beginning of the study, Model A was selected as the best model at 88\%. At the conclusion of the study, after users observed group fairness metrics, individual fairness metrics, and the exploration view, only 38\% selected Model A, with a plurality of the participants selecting the individually fair model (Model B). The exploration view allowed users to uncover more nuance in the dataset and discover the difficulty of determining the task of algorithmic fairness. 

\subsection{Exploration Strategies Overview}

\subsubsection{Strategies for Exploration: Group Fairness} 
\begin{figure*}[htbp]
\centering
\includegraphics[width=0.8\textwidth]{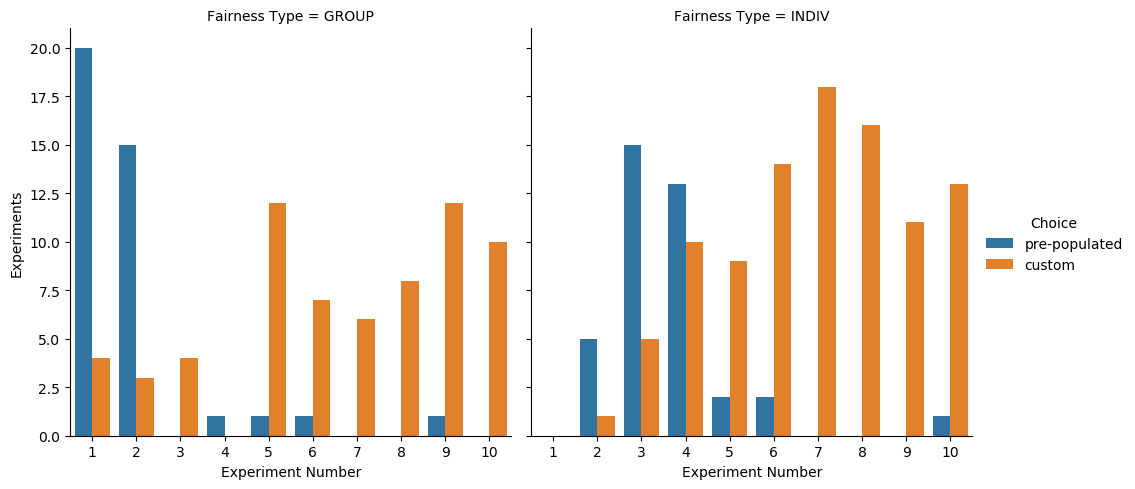}
\caption{Types of Experiments users completed during their 10 required experiments in the Exploration View.}
\label{fig:exresults}
\end{figure*}

We wanted to examine how practitioners in the study utilized the opportunity to input custom inputs and the strategies they took when doing so (\textbf{RQ2}). Our exploration page instructed users to run 10 experiments of their own. They could choose between group view metrics or individual metrics for each experiment. They could also select the pre-populated examples we had provided as a guide or choose to define free-form custom inputs themselves.  Figure \ref{fig:exresults} shows how users behaved for each of their experiments. Users initially started by using the guides provided and running group experiments with the pre-populated options given to them. In total, there were 62 custom group explorations that spanned topics like race (``asian'', ``black''), ethnicity (``korean'', ``mexican''), religion (``hindu'', ``christian'', ``mormon''), sexuality (``gay'', ``LGBT''), gender identity (``trans'', ``non-binary''), age (``old''), socio-economic class (``poor'', ``lower-class''), political identity (``communist'', ``democrat''), occupation (``sex worker''), and disability (``disabled'', ``handicapped'').

We also examined the cohesiveness of the groups. Users were able to provide multiple words to define a group. We wanted to see whether the words belonged to the same identity category, were part of different categories, or were groups that often were associated with one another. After analysis and coding of the groups users submitted for the custom group experiments, we characterize three strategies users took to identify groups (\textbf{RQ2a}): \textit{Same Identity Category but Different Identities}, \textit{Same Identity Category and similar identities} , \textit{Different Identity Category}. The categories along with the prevalence of each in the 62 explorations are listed below:

\begin{itemize}
    \item \textit{Same Identity Category but Different Identities} (81\%):: This occurs when users used words that were all related to the same category like race or gender, but the identities explored were different. For example, one user queried the words ``liberal, democrat, progressive, socialist'' as one group. While these are all political identities, they are political identities that appear on various sides of the politics spectrum. %
    \item \textit{Same Identity Category and Similar Identities} (14\%): We also observed users exploring the same identity categories with words that belonged to the same identity within that category. This means that the words used to define one group (i.e ``elderly, old people, old person'') can potentially be used interchangeably with one another. 
    \item \textit{Different Identity Categories} (5\%): This occurs when users chose to explore words that belonged to completely different identity categories as one group. For example, one user grouped the words ``BIPOC'' and ``transgender'' together as one group. Each of these words are descriptors for different categories of identities, where one can describe a person's race while another describes a person's gender identity. 
\end{itemize}

\subsubsection{Strategies for Exploration: Individual Fairness}
In total, we had 97 individual custom explorations. Custom explorations are when users opted not to use the pre-populated examples that we left for them and instead either altered the sentence or the identity tokens. Of the 97 examples, 54 (56\%) of the individual custom explorations kept one of the two (``yeah, and the \{WORD\} man is always innocent, even when he's guilty. right brock?'' and ``more evidence that trump's ban on \{WORD\} immigration is a good idea.'') pre-populated sentences in the examples and only altered the identity tokens.

The remaining sentences could be classified into two groups: Unambiguously Toxic (i.e ``I want to kill all the \{WORD\}'') and Unambiguously Non-Toxic (i.e ``Working \{WORD\} people all over the country are building a movement to take on corporate greed.''). Other themes that were selected in the sentences with which users experimented included current events (i.e \textit{Posted viral video in which two beautiful little \{WORD\} girls were blatantly rejected by a character they idolized}).\footnote{https://www.npr.org/2022/07/19/1112234153/sesame-place-apology-backlash-racism-rosita} Unlike the group exploration, many of the identity tokens that users wrote for the templates did not belong to the \textit{Same Identity Category and Similar Identities} categories, but rather used words that were notably different from one another within identity categories (i.e. ``white, black, hispanic''). When doing custom exploration with individual fairness metrics, users input both unambiguously toxic sentences and unambiguously non-toxic sentences utilizing identity tokens within the \textit{Same Identity Category and Similar Identities} category (\textbf{RQ2b}). 

\begin{quote}
    \textit{I aimed to test the model's ability to navigate nuance in language to see if honest, respectful discussion of potentially loaded topics would result in toxicity flags. My concern with anti-toxicity measures is that they have the potential to shut down constructive discourse about controversial topics.}  Participant \#12
\end{quote}

\subsection{Self-reported Reasons for Individual Exploration} 
We asked users to provide reasons for why they explored particular words in the custom view exploration. \bl{Two authors} coded each reason \bl{independently} through open coding  and generated the themes seen in Table \ref{tab:reasons2}. 
Users leveraged sentences seen in articles reporting on current events, racial categories (general and those who are historically discriminated against), personal experience, and ambiguous words to ``trick the model'' (\textbf{RQ2b}).

\begin{table*}

\centering
\begin{tabular}{p{2cm}|p{4cm}|p{3cm}| p{4cm}} 
\small
Reason &Definition&Words and Sentences Used &Example\\
\hline
\hline
Discriminated Groups&Sentences created included identity tokens of races/identities who have had a history of being discriminated against &gay, gays, Black, African American&	I was just trying to think of groups of people who are sometimes discriminated against.\\			
Malicious Content & Inclusive of hate speech and expletives & [redacted]	&Whether the phrase seemed to indicate malicious or disparaging intent		\\	
General Racial Categories & Sentences created included identity tokens of various races &Black, African American, Asian, Chinese, Japanese	& I think I was just trying to get a general sense by entering frequently used racial categories.	\\		
Personal Experience & Sentences were constructed based on an individual's personal experiences& [redacted] &	My own online experience. I have often seen obviously-toxic comments get through various filters, so I tried to simulate those kinds of comments\\			
Tricking the Model &Users tried to ``trick'' the model using words that may be toxic in some scenarios and non-toxic in others. & cheese, short &	I chose words that I thought would trick the model, like 'cheese', and ambiguous ones that might have negative connotations like 'short'\\			
Current Events	& Sentences were selected that reflected current events &Nichelle Nichols showed us the extraordinary power of \{bisexual,white,black\} women and paved the way for a better future for all women in media. &I picked words from recent news articles that I felt were neutral. \\			
Synonyms and Antonyms& Sentences were constructed in which the tokens explored were antonyms& white, black &I tried to use synonyms and antonyms and wanted to see how model A performs. \\

\hline
\end{tabular}

\caption{\label{tab:reasons2}Emergent themes from reasons why users selected custom inputs.}
\end{table*}%

\subsection{Subjective Preferences of Fairness Views and User Feedback}
We also asked participants to report their subjective preferences - to rank the four different views they were able to interact with in the Exploration Phase in order of which helped them the most to determine the fairness of the models. The rankings can be seen in Figure \ref{fig:subjective} with Pre-populated group view being ranked most highly, and Custom group view being ranked least. We asked participants to give reasons and insights behind the rankings. Despite providing examples, users expressed that they had trouble coming up with their own groups (Custom Group View) and sentences (Custom Individual View). Both pre-populated interactions were rated more highly than the custom views. Below, we include some of the feedback expressed by users for the different views:

\begin{figure*}[htbp]
\centering
\includegraphics[width=0.7\textwidth]{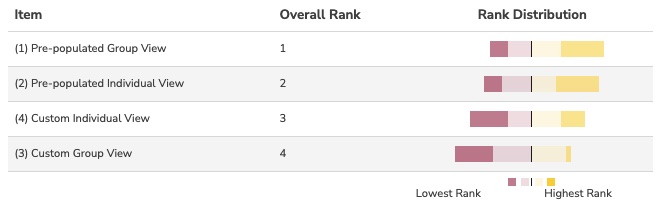}
\caption{Subjective preferences for the interaction view that helped the most to determine the fairness of the model. }
\label{fig:subjective}
\end{figure*}

\begin{quote}
    \textit{I found it very helpful to see the drops in accuracy in toxic and non-toxic comments for different models that mention minority groups.} \small{Participant \#8 , Pre-populated Group View} 
\end{quote}

\begin{quote}
    \textit{It was also cool to experiment with it using my own words but hard to think of some words fast enough. I think it might be useful to give some example types of subgroups beyond religion and ethnicity.}  \small{Participant \#74, Custom Group View} 
\end{quote}

\begin{quote}
    \textit{I had a hard time coming up with sentences. The one I tried weren't ranked as toxic by any model so it didn't really help.} \small{Participant \#2, Custom Individual View}
\end{quote}

\begin{quote}
    \textit{It was very useful to compare the models. It was easy to understand because it was binary (toxic vs non-toxic) compared to other numeric performance metrics, which is hard to define the threshold between good "enough" vs no} \small {Participant \#74, Pre-populated individual view} 
\end{quote}

We asked participants at the end of their task to reflect on the best/worst parts of their experience. We categorized the feedback into two groups: positive feedback and negative feedback. Of the positive feedback we received, users reported enjoying the customization, feeling empowered, and fully understanding the difficulty of the problem. Of the negative feedback, users reported feeling uncertainty about the words to use in the custom input and uncertainty about ``cost asymmetry of false positives vs. false negatives.''

\begin{quote}
\textit{It made me feel a little empowered like I might be helping those who are gonna dictate speech in the future that an AI model might not be enough to determine toxicity of a given sentence or grouping of words. } \small{Participant \#6}
\end{quote}

\section{Discussion}
In this section we provide design recommendations for building interfaces and tools for fairness exploration of text classifications. These recommendations are based on our findings on strategies participants in our study used to determine the fairness of a model, and the reasons behind those strategies.

\subsection{Different Views Lead to Different User Priorities and Different Model Selections: \bl{Group vs. Individual}}

Many users changed their decisions on their preferred model to deploy depending on view of the metrics with which they were presented (accuracy, group, individual, exploratory) (\textbf{RQ3}). We observed that the majority of the participants selected the ``regular'' model (Model A) when viewing performance metrics in Task 1, the group-fair model (Model C) when viewing group fairness metrics in Task 2, and the individually fair model (Model B) when viewing the individual fairness metrics in Task 3. In summary, the view designs, the models participants selected in the corresponding tasks, and the model training objectives were all aligned.

When identifying reasons why users made selections, we observed a similar shift in priorities. At the outset of the experiment, the majority of the users (88\%) selected Model A after seeing only the accuracy metrics, citing toxic accuracy as the reason for their selection. Upon seeing the group view and the toxic/non-toxic accuracy metrics for different minority groups, half of the participants selected Model C, citing non-toxic accuracy and the dangers of predicting text mentioning a minority group as toxic regardless of the content. After seeing examples of toxicity, majority of participants selected Model B (67\%) based on its resistance to prediction change when altering the identity tokens.  \bl{If the kind of fairness metrics presented to users can impact user perspectives, it is important for users to see both group and individual fairness. Our findings show different strategies used to explore group and individual fairness. The interface for group view allowed users to explore large amounts of data but gave users less flexibility by limiting them to identity tokens, whereas the individual view allowed users to explore the model output on one sentence at a time but gave them more flexibility to explore themes like personal experience or current events. There are benefits and limitations to both strategies and their effectiveness depends on the data being explored and the application of the model being evaluated. }

In line with prior work, we find that accuracy alone often leads to selection of a biased model, and providing perspectives on both group and individual fairness is important to allow users to make well-informed decisions
\cite{richardson2021towards}. We also emphasize the importance of developing a methodology for training models that can achieve group and individual fairness \emph{simultaneously} to satisfy a wider range of user priorities, which isn't the same as merely presenting a balanced accuracy metric. Currently, to the best of our knowledge, such methods are lacking and algorithmic fairness researchers tend to focus on \emph{either} group or individual fairness.

\subsection{Design Recommendations}

\bl{As language technologies and NLP become more widely deployed in various aspects of society, there are concerns about the harms they cause to various demographic groups. The focus of our work was on toxic text classification, but prior work has also revealed bias issues along demographic dimensions like race, sexuality, and gender through identity terms for other NLP tasks including question answering \cite{li2020unqovering}, relation extraction \cite{gaut2019towards}, occupation prediction \cite{de2019bias}, autocomplete generation \cite{sheng2019woman}, and machine translation \cite{stanovsky2019evaluating}. In this section, we provide design recommendations based on our findings. 
Many of our design recommendations generalize to a variety of NLP domains as well as to supervised learning broadly.}

\begin{itemize}
\item \textbf{Include Context} \bl{Because fairness can be so context dependent \cite{holstein2019improving}, contextual information can help practitioners better arrive at decisions. Our study is in the context of text classification, so context would include sentences/documents that appear before or after the document being evaluated for fairness. For NLP tasks, this could also include sentences that appear before/after the document being evaluated. For other domains, context could include all demographic information associated with an individual for loan applications \cite{nakao2022towards} and AI-driven decision making systems in healthcare \cite{chen2021algorithm}.} Showing additional context/metadata would be able to further assist the user in user decision-making. 
\item \textbf{\bl{Provide} Similarity Indicators} The group view of the data allowed users to explore accuracy metrics of subsets of the data that included different identity tokens. We found that participants used various strategies, either choosing groupings of words that belonged to the same identity category (black, African American) or groupings of words that belonged to different identity categories (Hindu, lesbian). These approaches yield different types of groups of data, however, given the dataset, it might not make sense to group identity tokens from different categories. For future interfaces that show group fairness in text classification, it could be useful for users to understand (1) how similar the sentences/documents that include the words they have chosen are to one another and (2) whether there is any overlap in the sentences based on the words they have selected. These indicators would give users additional guidance when defining the groups. \bl{Similarly, when comparing individual instances, participants made selections of both general racial categories and groups that are discriminated against. Seeing similarity indicators for identity tokens when viewing individual instances would give users more context. Our findings support prior work that explored practitioner perspectives of fairness in the context of loan applications. They found that participants liked the ability to see and compare individual applications and more information should be provided about the similarity of cases across decision boundaries \cite{nakao2022responsible}.}
\item \textbf{Include an Asymmetrical Counterfactual Metric} Prior work investigating fairness in text classification introduced the counterfactual fairness metric \cite{garg2019counterfactual} to generate counterfactuals with different identity tokens where text references a stereotype for one identity but not the other.
In the individual view examples, in which our participants explored sentences, we did not observe any sentences in which there would be asymmetrical counterfactuals for an identifier. Instead users tested either \textit{Unambiguously Non-Toxic} words or \textit{Unambiguously Toxic} words checking whether the model changed its classifications for different identifiers to determine whether a model is fair.
However, for practitioners and designers building tools for individual fairness exploration of text, including an asymmetrical counterfactual metric could potentially guide users to test those kinds of sentences that may be toxic for certain identity tokens and non-toxic for others. \bl{This metric would be helpful for other NLP tasks that exhibit bias along demographic dimensions like gender/race/religion/sexuality through identity terms.}
\item \textbf{Risk Associated with Overpredicting/Underpredicting} One theme that consistently emerged in the reasons users cited when making selections after seeing the accuracy view, was the risks of ``over-predicting'' toxic posts as opposed to ``under-predicting'' toxic posts. While a lot of studies have focused on evaluating with the F measure \cite{powers2020evaluation}, there are instances in which it makes sense to focus more on true negative rate (TNR) over true positive rate (TPR) (or vice versa). For example, when the cost and risk associated with a false positive is high, i.e. filtering out a legitimate/important email as spam \cite{kufandirimbwa2012spam}. Conversely, there are situations in which TPR is prioritized over TNR, for example in fraud detection or even disease detection \cite{vorobyev2022reducing}. 
It was clear that different participants held different perspectives on prioritizing non-toxic accuracy (TNR) or toxic accuracy (TPR) after seeing the accuracy view. Reasons for prioritizing non-toxic accuracy were to prevent penalizing content from ``good users'', while some people said in such a context, it is more important to overpredict toxic posts than underpredict toxic posts. Because there are various scenarios where overpredicting should be considered over underpredicting and vice versa, the entire pipeline must be considered. For example, once the data is classified, for the content classified falsely as toxic, are they simply flagged but remain in the data, or are they removed entirely? Conversely what happens to the content that is toxic but is not captured by the classifier? Is the population this content will be presented to at risk? Practitioners should be involved in the decision making of the outcome of the posts when deciding which model to employ, including all stakeholders involved.
\item \textbf{Provide \bl{Simulated-based approaches}}  \bl{In} our tool, users could create their own groups, sentences, and identity tokens for those sentences. However, when we asked users to rank their preferred interactions on the exploration page, both custom interactions were ranked less than the pre-populated examples. One user expressed, ``I had a hard time coming up with sentences. The one I tried weren't ranked as toxic by any model so it didn't really help.'' \bl{Prior work suggests to simulate conversational trajectories \cite{jain2018user} to support practitioners in their evaluation of fairness \cite {holstein2019improving}. Our findings support this notion, showing that users liked to have a few test examples before coming up with their own.} 
\item \textbf{Provide Additional Group Fairness Metrics} The high percentage of users (42\%) selecting Model A following the group view metrics (in Task 2) shows that just presenting users with group accuracy metrics does not sufficiently capture group fairness violations. Model A overpredicts toxicity when there are mentions of minority identities. For example, in Figure \ref{fig:group}, Model A's nontoxic accuracy is extremely low 59.5\%, while toxic accuracy is 94.2\%, demonstrating its tendency to overpredict toxicity for text with the corresponding group identity tokens (note also lower accuracy and balanced accuracy metrics). It is clear from the responses that participants who selected Model A were well-intentioned (focusing on capturing all toxic posts) but this selection comes at a cost of misclassifing text as toxic when it contains words referring to minorities. One way to mitigate this is to include additional group fairness metrics such as equalized odds violation \cite{hardt2016equality}. \bl{Similar problems arise in any supervised learning context, e.g., in criminal justice focusing only on the ability of a model to identify future recidivists might lead to preferring a model biased against black defendants \cite{angwin2016machine}.}
\item \textbf{Team Diversity} One of the emerging themes from our analysis of the kinds of token identities users selected in the exploration view was that users were using words from personal experience. This strategy - to explore words observed based on your personal experience - demonstrates the importance of diversity of teams, people with different backgrounds (sexuality, religion, race) who can explore various facets of the data and uncover potential bias. \bl{Recruiting diverse teams can be helpful in mitigating challenges due to ``blind spots'' and identifying which subpopulations to consider when considering fairness for different kinds of machine learning applications \cite{holstein2019improving}.} While not a design recommendation per se, many of the AI systems we create are a direct result of the data we collect, pre-process, feature engineer, and evaluate. Model evaluation by a group of diverse machine learning practitioners can potentially lead to selection of fairer models.

\end{itemize}

\section{Limitations}
\bl{While all our design recommendations may not necessarily generalize to all other domains of fairness perception, many of our design recommendations apply to NLP technologies that exhibit bias along demographic dimensions through identity terms.} Furthermore, we wanted to explore the kinds of strategies used by machine learning practitioners when interacting with a tool that shows both group and individual fairness. We provided a few examples as guidance and acknowledge that the pre-populated examples we provided may have impacted the kinds of custom examples users explored. However, in prior tasks leading up to the exploration page (Task 2, Task 3) we aimed to provide as many diverse examples as possible, that included different types of identities and sentences (sexuality, religion, ethnicity, gender identity). Future work can examine the impact of different pre-populated group and sentences examples on user strategies in a more controlled environment.

\bl{In the Group View task as well as our exploration task, we use group identifier keywords as a proxy for text content, which is a common way to find groups in text classification \cite{dixon2018measuring,park2018reducing,koh2021wilds,soares2022your}. However, we acknowledge that there may be toxic comments about groups of individuals without explicitly mentioning them \cite{borkan2019nuanced}. Building on prior work \cite{dixon2018measuring,garg2019counterfactual,yurochkin2021sensei,koh2021wilds,soares2022your}, we focus on the perceptions of the fairness of classification/misclassification of text inclusive of identity terms. In future work, we can examine alternative approaches to grouping text in which identities are implicitly referenced.}

\bl{One challenging aspect of this work is the subjective nature of user perception of toxicity. An individual’s perception of what is deemed offensive or toxic is influenced by not only the context of the text being presented to the individual but personal experiences, background and other individual factors \cite{zotti2019individual}. Prior work sheds light on differing individual toxicity perceptions in the context of online content moderation \cite{jiang2021understanding,marshall2021algorithmic,kumar2021designing}. It is important to study fairness perceptions, to see how HCI researchers can design interfaces and systems to account for both contextual diversity and the diversity of individuals who assess fairness. We believe our findings shed light on ways we can account for the subjectivity and contextual nature of fairness perception of toxicity.} We recognize that there is subjectivity in users' perceptions of toxicity, which may be influenced by personal background and other factors. While we analyzed the toxicity scores of the annotators, we did not investigate whether there was a consistent perception of toxicity among participants. This is an important issue that should be considered in future research. To provide a more accurate baseline, it may be necessary to conduct such an investigation with a larger and more diverse sample of participants.

Lastly, in this study we do not investigate the impact of viewing a certain type of fairness metric on the conclusion that a Machine Learning practitioner may draw about a Model. The initial tasks that users observed were meant as a tutorial before their experimentation stage. In future work, an experiment in which ordering effects are considered could be used to determine if a particular perspective of the data impacts the conclusions users may draw. Similarly, users could be given more freedom in determining how many experiments they perform before choosing a model in the Exploration View to understand disparity in explorations between practitioners and/or how many explorations might be considered sufficient in a given situation.

\section{Conclusion}
In this work, we investigate factors that influence a user's fairness decisions on various models when presented with different aspects and views of the data, strategies machine learning practitioners take to determine whether a model is fair, and the motivations behind those strategies. While other studies have examined user perspectives of fairness tools and indicators, our inquiry is the first to explore user interaction with fairness tools that examine text classification. We find that different views impact how users perceive the data and which metrics to prioritize (i.e. toxic accuracy/non-toxic accuracy/group fairness/individual fairness). We also identify different strategies users employ to determine the fairness of a model, forming groups of identities from similar categories or groups of identities from different categories (race, sexuality, gender). Based on our findings we are able to make recommendations for how to leverage front-end tooling to better assist users to employ strategies that will result in them forming a better picture of the data and model, ultimately making more informed decisions about a model's fairness and for how they prefer to interact with the tooling.

\balance{}

\bibliographystyle{ACM-Reference-Format}
\bibliography{sample}


\begin{thebibliography}{70}


\ifx \showCODEN    \undefined \def \showCODEN     #1{\unskip}     \fi
\ifx \showDOI      \undefined \def \showDOI       #1{#1}\fi
\ifx \showISBNx    \undefined \def \showISBNx     #1{\unskip}     \fi
\ifx \showISBNxiii \undefined \def \showISBNxiii  #1{\unskip}     \fi
\ifx \showISSN     \undefined \def \showISSN      #1{\unskip}     \fi
\ifx \showLCCN     \undefined \def \showLCCN      #1{\unskip}     \fi
\ifx \shownote     \undefined \def \shownote      #1{#1}          \fi
\ifx \showarticletitle \undefined \def \showarticletitle #1{#1}   \fi
\ifx \showURL      \undefined \def \showURL       {\relax}        \fi
\providecommand\bibfield[2]{#2}
\providecommand\bibinfo[2]{#2}
\providecommand\natexlab[1]{#1}
\providecommand\showeprint[2][]{arXiv:#2}

\bibitem[\protect\citeauthoryear{Agarwal, Beygelzimer, Dud{\'\i}k, Langford,
  and Wallach}{Agarwal et~al\mbox{.}}{2018}]%
        {agarwal2018reductions}
\bibfield{author}{\bibinfo{person}{Alekh Agarwal}, \bibinfo{person}{Alina
  Beygelzimer}, \bibinfo{person}{Miroslav Dud{\'\i}k}, \bibinfo{person}{John
  Langford}, {and} \bibinfo{person}{Hanna Wallach}.}
  \bibinfo{year}{2018}\natexlab{}.
\newblock \showarticletitle{A reductions approach to fair classification}. In
  \bibinfo{booktitle}{\emph{ICML}}.
\newblock


\bibitem[\protect\citeauthoryear{Angwin, Larson, Mattu, and Kirchner}{Angwin
  et~al\mbox{.}}{2016}]%
        {angwin2016machine}
\bibfield{author}{\bibinfo{person}{Julia Angwin}, \bibinfo{person}{Jeff
  Larson}, \bibinfo{person}{Surya Mattu}, {and} \bibinfo{person}{Lauren
  Kirchner}.} \bibinfo{year}{2016}\natexlab{}.
\newblock \showarticletitle{Machine bias}.
\newblock In \bibinfo{booktitle}{\emph{Ethics of Data and Analytics}}.
  \bibinfo{publisher}{Auerbach Publications}, \bibinfo{pages}{254--264}.
\newblock


\bibitem[\protect\citeauthoryear{Ball-Burack, Lee, Cobbe, and
  Singh}{Ball-Burack et~al\mbox{.}}{2021}]%
        {ball2021differential}
\bibfield{author}{\bibinfo{person}{Ari Ball-Burack}, \bibinfo{person}{Michelle
  Seng~Ah Lee}, \bibinfo{person}{Jennifer Cobbe}, {and}
  \bibinfo{person}{Jatinder Singh}.} \bibinfo{year}{2021}\natexlab{}.
\newblock \showarticletitle{Differential tweetment: Mitigating racial dialect
  bias in harmful tweet detection}. In \bibinfo{booktitle}{\emph{Proceedings of
  the 2021 ACM Conference on Fairness, Accountability, and Transparency}}.
  \bibinfo{pages}{116--128}.
\newblock


\bibitem[\protect\citeauthoryear{Barocas, Hardt, and Narayanan}{Barocas
  et~al\mbox{.}}{2019}]%
        {barocas2019Fairness}
\bibfield{author}{\bibinfo{person}{Solon Barocas}, \bibinfo{person}{Moritz
  Hardt}, {and} \bibinfo{person}{Arvind Narayanan}.}
  \bibinfo{year}{2019}\natexlab{}.
\newblock \bibinfo{booktitle}{\emph{Fairness and Machine Learning}}.
\newblock \bibinfo{publisher}{{fairmlbook.org}}.
\newblock


\bibitem[\protect\citeauthoryear{Bellamy, Dey, Hind, Hoffman, Houde, Kannan,
  Lohia, Martino, Mehta, Mojsilovi{\'c}, et~al\mbox{.}}{Bellamy
  et~al\mbox{.}}{2019}]%
        {bellamy2019ai}
\bibfield{author}{\bibinfo{person}{Rachel~KE Bellamy}, \bibinfo{person}{Kuntal
  Dey}, \bibinfo{person}{Michael Hind}, \bibinfo{person}{Samuel~C Hoffman},
  \bibinfo{person}{Stephanie Houde}, \bibinfo{person}{Kalapriya Kannan},
  \bibinfo{person}{Pranay Lohia}, \bibinfo{person}{Jacquelyn Martino},
  \bibinfo{person}{Sameep Mehta}, \bibinfo{person}{Aleksandra Mojsilovi{\'c}},
  {et~al\mbox{.}}} \bibinfo{year}{2019}\natexlab{}.
\newblock \showarticletitle{AI Fairness 360: An extensible toolkit for
  detecting and mitigating algorithmic bias}.
\newblock \bibinfo{journal}{\emph{IBM Journal of Research and Development}}
  \bibinfo{volume}{63}, \bibinfo{number}{4/5} (\bibinfo{year}{2019}),
  \bibinfo{pages}{4--1}.
\newblock


\bibitem[\protect\citeauthoryear{Berk, Heidari, Jabbari, Kearns, and Roth}{Berk
  et~al\mbox{.}}{2021}]%
        {berk2021fairness}
\bibfield{author}{\bibinfo{person}{Richard Berk}, \bibinfo{person}{Hoda
  Heidari}, \bibinfo{person}{Shahin Jabbari}, \bibinfo{person}{Michael Kearns},
  {and} \bibinfo{person}{Aaron Roth}.} \bibinfo{year}{2021}\natexlab{}.
\newblock \showarticletitle{Fairness in criminal justice risk assessments: The
  state of the art}.
\newblock \bibinfo{journal}{\emph{Sociological Methods \& Research}}
  \bibinfo{volume}{50}, \bibinfo{number}{1} (\bibinfo{year}{2021}),
  \bibinfo{pages}{3--44}.
\newblock


\bibitem[\protect\citeauthoryear{Bertrand and Mullainathan}{Bertrand and
  Mullainathan}{2004}]%
        {bertrand2004emily}
\bibfield{author}{\bibinfo{person}{Marianne Bertrand} {and}
  \bibinfo{person}{Sendhil Mullainathan}.} \bibinfo{year}{2004}\natexlab{}.
\newblock \showarticletitle{Are Emily and Greg more employable than Lakisha and
  Jamal? A field experiment on labor market discrimination}.
\newblock \bibinfo{journal}{\emph{American Economic Review}}
  \bibinfo{volume}{94}, \bibinfo{number}{4} (\bibinfo{year}{2004}),
  \bibinfo{pages}{991--1013}.
\newblock


\bibitem[\protect\citeauthoryear{Bird, Dud{\'\i}k, Edgar, Horn, Lutz, Milan,
  Sameki, Wallach, and Walker}{Bird et~al\mbox{.}}{2020}]%
        {bird2020fairlearn}
\bibfield{author}{\bibinfo{person}{Sarah Bird}, \bibinfo{person}{Miro
  Dud{\'\i}k}, \bibinfo{person}{Richard Edgar}, \bibinfo{person}{Brandon Horn},
  \bibinfo{person}{Roman Lutz}, \bibinfo{person}{Vanessa Milan},
  \bibinfo{person}{Mehrnoosh Sameki}, \bibinfo{person}{Hanna Wallach}, {and}
  \bibinfo{person}{Kathleen Walker}.} \bibinfo{year}{2020}\natexlab{}.
\newblock \showarticletitle{Fairlearn: A toolkit for assessing and improving
  fairness in AI}.
\newblock \bibinfo{journal}{\emph{Microsoft, Tech. Rep. MSR-TR-2020-32}}
  (\bibinfo{year}{2020}).
\newblock


\bibitem[\protect\citeauthoryear{Borkan, Dixon, Sorensen, Thain, and
  Vasserman}{Borkan et~al\mbox{.}}{2019}]%
        {borkan2019nuanced}
\bibfield{author}{\bibinfo{person}{Daniel Borkan}, \bibinfo{person}{Lucas
  Dixon}, \bibinfo{person}{Jeffrey Sorensen}, \bibinfo{person}{Nithum Thain},
  {and} \bibinfo{person}{Lucy Vasserman}.} \bibinfo{year}{2019}\natexlab{}.
\newblock \showarticletitle{Nuanced metrics for measuring unintended bias with
  real data for text classification}. In \bibinfo{booktitle}{\emph{Companion
  proceedings of the 2019 world wide web conference}}.
  \bibinfo{pages}{491--500}.
\newblock


\bibitem[\protect\citeauthoryear{Calmon, Wei, Vinzamuri, Natesan~Ramamurthy,
  and Varshney}{Calmon et~al\mbox{.}}{2017}]%
        {calmon2017optimized}
\bibfield{author}{\bibinfo{person}{Flavio Calmon}, \bibinfo{person}{Dennis
  Wei}, \bibinfo{person}{Bhanukiran Vinzamuri}, \bibinfo{person}{Karthikeyan
  Natesan~Ramamurthy}, {and} \bibinfo{person}{Kush~R Varshney}.}
  \bibinfo{year}{2017}\natexlab{}.
\newblock \showarticletitle{Optimized pre-processing for discrimination
  prevention}.
\newblock \bibinfo{journal}{\emph{Advances in neural information processing
  systems}}  \bibinfo{volume}{30} (\bibinfo{year}{2017}).
\newblock


\bibitem[\protect\citeauthoryear{Chen, Chen, Lipkova, Wang, Williamson, Lu,
  Sahai, and Mahmood}{Chen et~al\mbox{.}}{2021}]%
        {chen2021algorithm}
\bibfield{author}{\bibinfo{person}{Richard~J Chen}, \bibinfo{person}{Tiffany~Y
  Chen}, \bibinfo{person}{Jana Lipkova}, \bibinfo{person}{Judy~J Wang},
  \bibinfo{person}{Drew~FK Williamson}, \bibinfo{person}{Ming~Y Lu},
  \bibinfo{person}{Sharifa Sahai}, {and} \bibinfo{person}{Faisal Mahmood}.}
  \bibinfo{year}{2021}\natexlab{}.
\newblock \showarticletitle{Algorithm fairness in ai for medicine and
  healthcare}.
\newblock \bibinfo{journal}{\emph{arXiv preprint arXiv:2110.00603}}
  (\bibinfo{year}{2021}).
\newblock


\bibitem[\protect\citeauthoryear{Chouldechova}{Chouldechova}{2017}]%
        {chouldechova2017fair}
\bibfield{author}{\bibinfo{person}{Alexandra Chouldechova}.}
  \bibinfo{year}{2017}\natexlab{}.
\newblock \showarticletitle{Fair prediction with disparate impact: A study of
  bias in recidivism prediction instruments}.
\newblock \bibinfo{journal}{\emph{Big data}} \bibinfo{volume}{5},
  \bibinfo{number}{2} (\bibinfo{year}{2017}), \bibinfo{pages}{153--163}.
\newblock


\bibitem[\protect\citeauthoryear{Chouldechova and Roth}{Chouldechova and
  Roth}{2020}]%
        {chouldechova2020snapshot}
\bibfield{author}{\bibinfo{person}{Alexandra Chouldechova} {and}
  \bibinfo{person}{Aaron Roth}.} \bibinfo{year}{2020}\natexlab{}.
\newblock \showarticletitle{A snapshot of the frontiers of fairness in machine
  learning}.
\newblock \bibinfo{journal}{\emph{Commun. ACM}} \bibinfo{volume}{63},
  \bibinfo{number}{5} (\bibinfo{year}{2020}), \bibinfo{pages}{82--89}.
\newblock


\bibitem[\protect\citeauthoryear{Cotter, Jiang, and Sridharan}{Cotter
  et~al\mbox{.}}{2019}]%
        {cotter2019two}
\bibfield{author}{\bibinfo{person}{Andrew Cotter}, \bibinfo{person}{Heinrich
  Jiang}, {and} \bibinfo{person}{Karthik Sridharan}.}
  \bibinfo{year}{2019}\natexlab{}.
\newblock \showarticletitle{Two-player games for efficient non-convex
  constrained optimization}. In \bibinfo{booktitle}{\emph{Algorithmic Learning
  Theory}}. PMLR, \bibinfo{pages}{300--332}.
\newblock


\bibitem[\protect\citeauthoryear{Crawford}{Crawford}{2017}]%
        {crawford2017trouble}
\bibfield{author}{\bibinfo{person}{Kate Crawford}.}
  \bibinfo{year}{2017}\natexlab{}.
\newblock \showarticletitle{The trouble with bias. keynote at neurips}.
\newblock  (\bibinfo{year}{2017}).
\newblock


\bibitem[\protect\citeauthoryear{De-Arteaga, Romanov, Wallach, Chayes, Borgs,
  Chouldechova, Geyik, Kenthapadi, and Kalai}{De-Arteaga et~al\mbox{.}}{2019}]%
        {de2019bias}
\bibfield{author}{\bibinfo{person}{Maria De-Arteaga}, \bibinfo{person}{Alexey
  Romanov}, \bibinfo{person}{Hanna Wallach}, \bibinfo{person}{Jennifer Chayes},
  \bibinfo{person}{Christian Borgs}, \bibinfo{person}{Alexandra Chouldechova},
  \bibinfo{person}{Sahin Geyik}, \bibinfo{person}{Krishnaram Kenthapadi}, {and}
  \bibinfo{person}{Adam~Tauman Kalai}.} \bibinfo{year}{2019}\natexlab{}.
\newblock \showarticletitle{Bias in bios: A case study of semantic
  representation bias in a high-stakes setting}. In
  \bibinfo{booktitle}{\emph{proceedings of the Conference on Fairness,
  Accountability, and Transparency}}. \bibinfo{pages}{120--128}.
\newblock


\bibitem[\protect\citeauthoryear{Deng, Nagireddy, Lee, Singh, Wu, Holstein, and
  Zhu}{Deng et~al\mbox{.}}{2022}]%
        {10.1145/3531146.3533113}
\bibfield{author}{\bibinfo{person}{Wesley~Hanwen Deng}, \bibinfo{person}{Manish
  Nagireddy}, \bibinfo{person}{Michelle Seng~Ah Lee}, \bibinfo{person}{Jatinder
  Singh}, \bibinfo{person}{Zhiwei~Steven Wu}, \bibinfo{person}{Kenneth
  Holstein}, {and} \bibinfo{person}{Haiyi Zhu}.}
  \bibinfo{year}{2022}\natexlab{}.
\newblock \showarticletitle{Exploring How Machine Learning Practitioners (Try
  To) Use Fairness Toolkits}. In \bibinfo{booktitle}{\emph{2022 ACM Conference
  on Fairness, Accountability, and Transparency}} (Seoul, Republic of Korea)
  \emph{(\bibinfo{series}{FAccT '22})}. \bibinfo{publisher}{Association for
  Computing Machinery}, \bibinfo{address}{New York, NY, USA},
  \bibinfo{pages}{473–484}.
\newblock
\showISBNx{9781450393522}
\urldef\tempurl%
\url{https://doi.org/10.1145/3531146.3533113}
\showDOI{\tempurl}


\bibitem[\protect\citeauthoryear{Dev, Sheng, Zhao, Amstutz, Sun, Hou,
  Sanseverino, Kim, Nishi, Peng, et~al\mbox{.}}{Dev et~al\mbox{.}}{2022}]%
        {dev2022measures}
\bibfield{author}{\bibinfo{person}{Sunipa Dev}, \bibinfo{person}{Emily Sheng},
  \bibinfo{person}{Jieyu Zhao}, \bibinfo{person}{Aubrie Amstutz},
  \bibinfo{person}{Jiao Sun}, \bibinfo{person}{Yu Hou}, \bibinfo{person}{Mattie
  Sanseverino}, \bibinfo{person}{Jiin Kim}, \bibinfo{person}{Akihiro Nishi},
  \bibinfo{person}{Nanyun Peng}, {et~al\mbox{.}}}
  \bibinfo{year}{2022}\natexlab{}.
\newblock \showarticletitle{On Measures of Biases and Harms in NLP}. In
  \bibinfo{booktitle}{\emph{Findings of the Association for Computational
  Linguistics: AACL-IJCNLP 2022}}. \bibinfo{pages}{246--267}.
\newblock


\bibitem[\protect\citeauthoryear{Devlin, Chang, Lee, and Toutanova}{Devlin
  et~al\mbox{.}}{2018}]%
        {devlin2018bert}
\bibfield{author}{\bibinfo{person}{Jacob Devlin}, \bibinfo{person}{Ming-Wei
  Chang}, \bibinfo{person}{Kenton Lee}, {and} \bibinfo{person}{Kristina
  Toutanova}.} \bibinfo{year}{2018}\natexlab{}.
\newblock \showarticletitle{Bert: Pre-training of deep bidirectional
  transformers for language understanding}.
\newblock \bibinfo{journal}{\emph{arXiv preprint arXiv:1810.04805}}
  (\bibinfo{year}{2018}).
\newblock


\bibitem[\protect\citeauthoryear{Dixon, Li, Sorensen, Thain, and
  Vasserman}{Dixon et~al\mbox{.}}{2018}]%
        {dixon2018measuring}
\bibfield{author}{\bibinfo{person}{Lucas Dixon}, \bibinfo{person}{John Li},
  \bibinfo{person}{Jeffrey Sorensen}, \bibinfo{person}{Nithum Thain}, {and}
  \bibinfo{person}{Lucy Vasserman}.} \bibinfo{year}{2018}\natexlab{}.
\newblock \showarticletitle{Measuring and mitigating unintended bias in text
  classification}. In \bibinfo{booktitle}{\emph{Proceedings of the 2018
  AAAI/ACM Conference on AI, Ethics, and Society}}. \bibinfo{pages}{67--73}.
\newblock


\bibitem[\protect\citeauthoryear{Dwork, Hardt, Pitassi, Reingold, and
  Zemel}{Dwork et~al\mbox{.}}{2012}]%
        {dwork2012fairness}
\bibfield{author}{\bibinfo{person}{Cynthia Dwork}, \bibinfo{person}{Moritz
  Hardt}, \bibinfo{person}{Toniann Pitassi}, \bibinfo{person}{Omer Reingold},
  {and} \bibinfo{person}{Richard Zemel}.} \bibinfo{year}{2012}\natexlab{}.
\newblock \showarticletitle{Fairness through awareness}. In
  \bibinfo{booktitle}{\emph{ITCS}}.
\newblock


\bibitem[\protect\citeauthoryear{Fleisher}{Fleisher}{2021}]%
        {fleisher2021s}
\bibfield{author}{\bibinfo{person}{Will Fleisher}.}
  \bibinfo{year}{2021}\natexlab{}.
\newblock \showarticletitle{What's Fair about Individual Fairness?}. In
  \bibinfo{booktitle}{\emph{Proceedings of the 2021 AAAI/ACM Conference on AI,
  Ethics, and Society}}. \bibinfo{pages}{480--490}.
\newblock


\bibitem[\protect\citeauthoryear{Friedler, Scheidegger, Venkatasubramanian,
  Choudhary, Hamilton, and Roth}{Friedler et~al\mbox{.}}{2019}]%
        {friedler2019comparative}
\bibfield{author}{\bibinfo{person}{Sorelle~A Friedler}, \bibinfo{person}{Carlos
  Scheidegger}, \bibinfo{person}{Suresh Venkatasubramanian},
  \bibinfo{person}{Sonam Choudhary}, \bibinfo{person}{Evan~P Hamilton}, {and}
  \bibinfo{person}{Derek Roth}.} \bibinfo{year}{2019}\natexlab{}.
\newblock \showarticletitle{A comparative study of fairness-enhancing
  interventions in machine learning}. In \bibinfo{booktitle}{\emph{Proceedings
  of the conference on fairness, accountability, and transparency}}.
  \bibinfo{pages}{329--338}.
\newblock


\bibitem[\protect\citeauthoryear{Garg, Perot, Limtiaco, Taly, Chi, and
  Beutel}{Garg et~al\mbox{.}}{2019}]%
        {garg2019counterfactual}
\bibfield{author}{\bibinfo{person}{Sahaj Garg}, \bibinfo{person}{Vincent
  Perot}, \bibinfo{person}{Nicole Limtiaco}, \bibinfo{person}{Ankur Taly},
  \bibinfo{person}{Ed~H Chi}, {and} \bibinfo{person}{Alex Beutel}.}
  \bibinfo{year}{2019}\natexlab{}.
\newblock \showarticletitle{Counterfactual fairness in text classification
  through robustness}. In \bibinfo{booktitle}{\emph{Proceedings of the 2019
  AAAI/ACM Conference on AI, Ethics, and Society}}. \bibinfo{pages}{219--226}.
\newblock


\bibitem[\protect\citeauthoryear{Gaut, Sun, Tang, Huang, Qian, ElSherief, Zhao,
  Mirza, Belding, Chang, et~al\mbox{.}}{Gaut et~al\mbox{.}}{2019}]%
        {gaut2019towards}
\bibfield{author}{\bibinfo{person}{Andrew Gaut}, \bibinfo{person}{Tony Sun},
  \bibinfo{person}{Shirlyn Tang}, \bibinfo{person}{Yuxin Huang},
  \bibinfo{person}{Jing Qian}, \bibinfo{person}{Mai ElSherief},
  \bibinfo{person}{Jieyu Zhao}, \bibinfo{person}{Diba Mirza},
  \bibinfo{person}{Elizabeth Belding}, \bibinfo{person}{Kai-Wei Chang},
  {et~al\mbox{.}}} \bibinfo{year}{2019}\natexlab{}.
\newblock \showarticletitle{Towards understanding gender bias in relation
  extraction}.
\newblock \bibinfo{journal}{\emph{arXiv preprint arXiv:1911.03642}}
  (\bibinfo{year}{2019}).
\newblock


\bibitem[\protect\citeauthoryear{Hardt, Price, and Srebro}{Hardt
  et~al\mbox{.}}{2016}]%
        {hardt2016equality}
\bibfield{author}{\bibinfo{person}{Moritz Hardt}, \bibinfo{person}{Eric Price},
  {and} \bibinfo{person}{Nati Srebro}.} \bibinfo{year}{2016}\natexlab{}.
\newblock \showarticletitle{Equality of opportunity in supervised learning}.
\newblock \bibinfo{journal}{\emph{Advances in neural information processing
  systems}}  \bibinfo{volume}{29} (\bibinfo{year}{2016}).
\newblock


\bibitem[\protect\citeauthoryear{Hofstede}{Hofstede}{2011}]%
        {hofstede2011dimensionalizing}
\bibfield{author}{\bibinfo{person}{Geert Hofstede}.}
  \bibinfo{year}{2011}\natexlab{}.
\newblock \showarticletitle{Dimensionalizing cultures: The Hofstede model in
  context}.
\newblock \bibinfo{journal}{\emph{Online readings in psychology and culture}}
  \bibinfo{volume}{2}, \bibinfo{number}{1} (\bibinfo{year}{2011}),
  \bibinfo{pages}{2307--0919}.
\newblock


\bibitem[\protect\citeauthoryear{Holstein, Wortman~Vaughan, Daum{\'e}~III,
  Dudik, and Wallach}{Holstein et~al\mbox{.}}{2019}]%
        {holstein2019improving}
\bibfield{author}{\bibinfo{person}{Kenneth Holstein}, \bibinfo{person}{Jennifer
  Wortman~Vaughan}, \bibinfo{person}{Hal Daum{\'e}~III}, \bibinfo{person}{Miro
  Dudik}, {and} \bibinfo{person}{Hanna Wallach}.}
  \bibinfo{year}{2019}\natexlab{}.
\newblock \showarticletitle{Improving fairness in machine learning systems:
  What do industry practitioners need?}. In
  \bibinfo{booktitle}{\emph{Proceedings of the 2019 CHI conference on human
  factors in computing systems}}. \bibinfo{pages}{1--16}.
\newblock


\bibitem[\protect\citeauthoryear{Ilvento}{Ilvento}{2019}]%
        {ilvento2019metric}
\bibfield{author}{\bibinfo{person}{Christina Ilvento}.}
  \bibinfo{year}{2019}\natexlab{}.
\newblock \showarticletitle{Metric learning for individual fairness}.
\newblock \bibinfo{journal}{\emph{arXiv preprint arXiv:1906.00250}}
  (\bibinfo{year}{2019}).
\newblock


\bibitem[\protect\citeauthoryear{Jain, Pecune, Matsuyama, and Cassell}{Jain
  et~al\mbox{.}}{2018}]%
        {jain2018user}
\bibfield{author}{\bibinfo{person}{Alankar Jain}, \bibinfo{person}{Florian
  Pecune}, \bibinfo{person}{Yoichi Matsuyama}, {and} \bibinfo{person}{Justine
  Cassell}.} \bibinfo{year}{2018}\natexlab{}.
\newblock \showarticletitle{A user simulator architecture for socially-aware
  conversational agents}. In \bibinfo{booktitle}{\emph{Proceedings of the 18th
  International Conference on Intelligent Virtual Agents}}.
  \bibinfo{pages}{133--140}.
\newblock


\bibitem[\protect\citeauthoryear{Jiang, Scheuerman, Fiesler, and
  Brubaker}{Jiang et~al\mbox{.}}{2021}]%
        {jiang2021understanding}
\bibfield{author}{\bibinfo{person}{Jialun~Aaron Jiang},
  \bibinfo{person}{Morgan~Klaus Scheuerman}, \bibinfo{person}{Casey Fiesler},
  {and} \bibinfo{person}{Jed~R Brubaker}.} \bibinfo{year}{2021}\natexlab{}.
\newblock \showarticletitle{Understanding international perceptions of the
  severity of harmful content online}.
\newblock \bibinfo{journal}{\emph{PloS one}} \bibinfo{volume}{16},
  \bibinfo{number}{8} (\bibinfo{year}{2021}), \bibinfo{pages}{e0256762}.
\newblock


\bibitem[\protect\citeauthoryear{John-Mathews, Cardon, and
  Balagu{\'e}}{John-Mathews et~al\mbox{.}}{2022}]%
        {john2022reality}
\bibfield{author}{\bibinfo{person}{Jean-Marie John-Mathews},
  \bibinfo{person}{Dominique Cardon}, {and} \bibinfo{person}{Christine
  Balagu{\'e}}.} \bibinfo{year}{2022}\natexlab{}.
\newblock \showarticletitle{From reality to world. A critical perspective on AI
  fairness}.
\newblock \bibinfo{journal}{\emph{Journal of Business Ethics}}
  (\bibinfo{year}{2022}), \bibinfo{pages}{1--15}.
\newblock


\bibitem[\protect\citeauthoryear{Kleinberg, Mullainathan, and
  Raghavan}{Kleinberg et~al\mbox{.}}{2016}]%
        {kleinberg2016inherent}
\bibfield{author}{\bibinfo{person}{Jon Kleinberg}, \bibinfo{person}{Sendhil
  Mullainathan}, {and} \bibinfo{person}{Manish Raghavan}.}
  \bibinfo{year}{2016}\natexlab{}.
\newblock \showarticletitle{Inherent trade-offs in the fair determination of
  risk scores}.
\newblock \bibinfo{journal}{\emph{arXiv preprint arXiv:1609.05807}}
  (\bibinfo{year}{2016}).
\newblock


\bibitem[\protect\citeauthoryear{Koh, Sagawa, Marklund, Xie, Zhang,
  Balsubramani, Hu, Yasunaga, Phillips, Gao, et~al\mbox{.}}{Koh
  et~al\mbox{.}}{2021}]%
        {koh2021wilds}
\bibfield{author}{\bibinfo{person}{Pang~Wei Koh}, \bibinfo{person}{Shiori
  Sagawa}, \bibinfo{person}{Henrik Marklund}, \bibinfo{person}{Sang~Michael
  Xie}, \bibinfo{person}{Marvin Zhang}, \bibinfo{person}{Akshay Balsubramani},
  \bibinfo{person}{Weihua Hu}, \bibinfo{person}{Michihiro Yasunaga},
  \bibinfo{person}{Richard~Lanas Phillips}, \bibinfo{person}{Irena Gao},
  {et~al\mbox{.}}} \bibinfo{year}{2021}\natexlab{}.
\newblock \showarticletitle{Wilds: A benchmark of in-the-wild distribution
  shifts}. In \bibinfo{booktitle}{\emph{International Conference on Machine
  Learning}}. PMLR, \bibinfo{pages}{5637--5664}.
\newblock


\bibitem[\protect\citeauthoryear{Krishnapriya, Albiero, Vangara, King, and
  Bowyer}{Krishnapriya et~al\mbox{.}}{2020}]%
        {krishnapriya2020issues}
\bibfield{author}{\bibinfo{person}{KS Krishnapriya},
  \bibinfo{person}{V{\'\i}tor Albiero}, \bibinfo{person}{Kushal Vangara},
  \bibinfo{person}{Michael~C King}, {and} \bibinfo{person}{Kevin~W Bowyer}.}
  \bibinfo{year}{2020}\natexlab{}.
\newblock \showarticletitle{Issues related to face recognition accuracy varying
  based on race and skin tone}.
\newblock \bibinfo{journal}{\emph{IEEE Transactions on Technology and Society}}
  \bibinfo{volume}{1}, \bibinfo{number}{1} (\bibinfo{year}{2020}),
  \bibinfo{pages}{8--20}.
\newblock


\bibitem[\protect\citeauthoryear{Kufandirimbwa and Gotora}{Kufandirimbwa and
  Gotora}{2012}]%
        {kufandirimbwa2012spam}
\bibfield{author}{\bibinfo{person}{Owen Kufandirimbwa} {and}
  \bibinfo{person}{Richard Gotora}.} \bibinfo{year}{2012}\natexlab{}.
\newblock \showarticletitle{Spam detection using artificial neural networks
  (perceptron learning rule)}.
\newblock \bibinfo{journal}{\emph{Online Journal of Physical and Environmental
  Science Research}} \bibinfo{volume}{1}, \bibinfo{number}{2}
  (\bibinfo{year}{2012}), \bibinfo{pages}{22--29}.
\newblock


\bibitem[\protect\citeauthoryear{Kumar, Kelley, Consolvo, Mason, Bursztein,
  Durumeric, Thomas, and Bailey}{Kumar et~al\mbox{.}}{2021}]%
        {kumar2021designing}
\bibfield{author}{\bibinfo{person}{Deepak Kumar}, \bibinfo{person}{Patrick~Gage
  Kelley}, \bibinfo{person}{Sunny Consolvo}, \bibinfo{person}{Joshua Mason},
  \bibinfo{person}{Elie Bursztein}, \bibinfo{person}{Zakir Durumeric},
  \bibinfo{person}{Kurt Thomas}, {and} \bibinfo{person}{Michael Bailey}.}
  \bibinfo{year}{2021}\natexlab{}.
\newblock \showarticletitle{Designing Toxic Content Classification for a
  Diversity of Perspectives.}. In \bibinfo{booktitle}{\emph{SOUPS@ USENIX
  Security Symposium}}. \bibinfo{pages}{299--318}.
\newblock


\bibitem[\protect\citeauthoryear{Li, Khot, Khashabi, Sabharwal, and
  Srikumar}{Li et~al\mbox{.}}{2020}]%
        {li2020unqovering}
\bibfield{author}{\bibinfo{person}{Tao Li}, \bibinfo{person}{Tushar Khot},
  \bibinfo{person}{Daniel Khashabi}, \bibinfo{person}{Ashish Sabharwal}, {and}
  \bibinfo{person}{Vivek Srikumar}.} \bibinfo{year}{2020}\natexlab{}.
\newblock \showarticletitle{UNQOVERing stereotyping biases via underspecified
  questions}.
\newblock \bibinfo{journal}{\emph{arXiv preprint arXiv:2010.02428}}
  (\bibinfo{year}{2020}).
\newblock


\bibitem[\protect\citeauthoryear{Madaio, Egede, Subramonyam, Wortman~Vaughan,
  and Wallach}{Madaio et~al\mbox{.}}{2022}]%
        {10.1145/3512899}
\bibfield{author}{\bibinfo{person}{Michael Madaio}, \bibinfo{person}{Lisa
  Egede}, \bibinfo{person}{Hariharan Subramonyam}, \bibinfo{person}{Jennifer
  Wortman~Vaughan}, {and} \bibinfo{person}{Hanna Wallach}.}
  \bibinfo{year}{2022}\natexlab{}.
\newblock \showarticletitle{Assessing the Fairness of AI Systems: AI
  Practitioners' Processes, Challenges, and Needs for Support}.
\newblock \bibinfo{journal}{\emph{Proc. ACM Hum.-Comput. Interact.}}
  \bibinfo{volume}{6}, \bibinfo{number}{CSCW1}, Article \bibinfo{articleno}{52}
  (\bibinfo{date}{apr} \bibinfo{year}{2022}), \bibinfo{numpages}{26}~pages.
\newblock
\urldef\tempurl%
\url{https://doi.org/10.1145/3512899}
\showDOI{\tempurl}


\bibitem[\protect\citeauthoryear{Madaio, Stark, Wortman~Vaughan, and
  Wallach}{Madaio et~al\mbox{.}}{2020}]%
        {madaio2020co}
\bibfield{author}{\bibinfo{person}{Michael~A Madaio}, \bibinfo{person}{Luke
  Stark}, \bibinfo{person}{Jennifer Wortman~Vaughan}, {and}
  \bibinfo{person}{Hanna Wallach}.} \bibinfo{year}{2020}\natexlab{}.
\newblock \showarticletitle{Co-designing checklists to understand
  organizational challenges and opportunities around fairness in AI}. In
  \bibinfo{booktitle}{\emph{Proceedings of the 2020 CHI Conference on Human
  Factors in Computing Systems}}. \bibinfo{pages}{1--14}.
\newblock


\bibitem[\protect\citeauthoryear{Maity, Xue, Yurochkin, and Sun}{Maity
  et~al\mbox{.}}{2021}]%
        {maity2021statistical}
\bibfield{author}{\bibinfo{person}{Subha Maity}, \bibinfo{person}{Songkai Xue},
  \bibinfo{person}{Mikhail Yurochkin}, {and} \bibinfo{person}{Yuekai Sun}.}
  \bibinfo{year}{2021}\natexlab{}.
\newblock \showarticletitle{Statistical inference for individual fairness}.
\newblock \bibinfo{journal}{\emph{arXiv preprint arXiv:2103.16714}}
  (\bibinfo{year}{2021}).
\newblock


\bibitem[\protect\citeauthoryear{Makhortykh, Urman, and Ulloa}{Makhortykh
  et~al\mbox{.}}{2021}]%
        {makhortykh2021detecting}
\bibfield{author}{\bibinfo{person}{Mykola Makhortykh},
  \bibinfo{person}{Aleksandra Urman}, {and} \bibinfo{person}{Roberto Ulloa}.}
  \bibinfo{year}{2021}\natexlab{}.
\newblock \showarticletitle{Detecting race and gender bias in visual
  representation of AI on web search engines}. In
  \bibinfo{booktitle}{\emph{International Workshop on Algorithmic Bias in
  Search and Recommendation}}. Springer, \bibinfo{pages}{36--50}.
\newblock


\bibitem[\protect\citeauthoryear{Mallari, Inkpen, Johns, Tan, Ramesh, and
  Kamar}{Mallari et~al\mbox{.}}{2020}]%
        {mallari2020look}
\bibfield{author}{\bibinfo{person}{Keri Mallari}, \bibinfo{person}{Kori
  Inkpen}, \bibinfo{person}{Paul Johns}, \bibinfo{person}{Sarah Tan},
  \bibinfo{person}{Divya Ramesh}, {and} \bibinfo{person}{Ece Kamar}.}
  \bibinfo{year}{2020}\natexlab{}.
\newblock \showarticletitle{Do i look like a criminal? examining how race
  presentation impacts human judgement of recidivism}. In
  \bibinfo{booktitle}{\emph{Proceedings of the 2020 Chi conference on human
  factors in computing systems}}. \bibinfo{pages}{1--13}.
\newblock


\bibitem[\protect\citeauthoryear{Marshall}{Marshall}{2021}]%
        {marshall2021algorithmic}
\bibfield{author}{\bibinfo{person}{Brandeis Marshall}.}
  \bibinfo{year}{2021}\natexlab{}.
\newblock \showarticletitle{Algorithmic misogynoir in content moderation
  practice}.
\newblock \bibinfo{journal}{\emph{Heinrich-B{\"o}ll-Stiftung European Union}}
  (\bibinfo{year}{2021}).
\newblock


\bibitem[\protect\citeauthoryear{Mukherjee, Yurochkin, Banerjee, and
  Sun}{Mukherjee et~al\mbox{.}}{2020}]%
        {mukherjee2020two}
\bibfield{author}{\bibinfo{person}{Debarghya Mukherjee},
  \bibinfo{person}{Mikhail Yurochkin}, \bibinfo{person}{Moulinath Banerjee},
  {and} \bibinfo{person}{Yuekai Sun}.} \bibinfo{year}{2020}\natexlab{}.
\newblock \showarticletitle{Two simple ways to learn individual fairness
  metrics from data}. In \bibinfo{booktitle}{\emph{ICML}}.
\newblock


\bibitem[\protect\citeauthoryear{Nakao, Strappelli, Stumpf, Naseer, Regoli, and
  Gamba}{Nakao et~al\mbox{.}}{2022a}]%
        {nakao2022responsible}
\bibfield{author}{\bibinfo{person}{Yuri Nakao}, \bibinfo{person}{Lorenzo
  Strappelli}, \bibinfo{person}{Simone Stumpf}, \bibinfo{person}{Aisha Naseer},
  \bibinfo{person}{Daniele Regoli}, {and} \bibinfo{person}{Giulia~Del Gamba}.}
  \bibinfo{year}{2022}\natexlab{a}.
\newblock \showarticletitle{Towards Responsible AI: A Design Space Exploration
  of Human-Centered Artificial Intelligence User Interfaces to Investigate
  Fairness}.
\newblock \bibinfo{journal}{\emph{International Journal of Human--Computer
  Interaction}} (\bibinfo{year}{2022}), \bibinfo{pages}{1--27}.
\newblock


\bibitem[\protect\citeauthoryear{Nakao, Stumpf, Ahmed, Naseer, and
  Strappelli}{Nakao et~al\mbox{.}}{2022b}]%
        {nakao2022towards}
\bibfield{author}{\bibinfo{person}{Yuri Nakao}, \bibinfo{person}{Simone
  Stumpf}, \bibinfo{person}{Subeida Ahmed}, \bibinfo{person}{Aisha Naseer},
  {and} \bibinfo{person}{Lorenzo Strappelli}.}
  \bibinfo{year}{2022}\natexlab{b}.
\newblock \showarticletitle{Towards Involving End-users in Interactive
  Human-in-the-loop AI Fairness}.
\newblock \bibinfo{journal}{\emph{ACM Transactions on Interactive Intelligent
  Systems (TiiS)}} (\bibinfo{year}{2022}).
\newblock


\bibitem[\protect\citeauthoryear{Park, Shin, and Fung}{Park
  et~al\mbox{.}}{2018}]%
        {park2018reducing}
\bibfield{author}{\bibinfo{person}{Ji~Ho Park}, \bibinfo{person}{Jamin Shin},
  {and} \bibinfo{person}{Pascale Fung}.} \bibinfo{year}{2018}\natexlab{}.
\newblock \showarticletitle{Reducing Gender Bias in Abusive Language
  Detection}. In \bibinfo{booktitle}{\emph{Proceedings of the 2018 Conference
  on Empirical Methods in Natural Language Processing, EMNLP 2018}}.
\newblock


\bibitem[\protect\citeauthoryear{Petersen, Mukherjee, Sun, and
  Yurochkin}{Petersen et~al\mbox{.}}{2021}]%
        {petersen2021post}
\bibfield{author}{\bibinfo{person}{Felix Petersen}, \bibinfo{person}{Debarghya
  Mukherjee}, \bibinfo{person}{Yuekai Sun}, {and} \bibinfo{person}{Mikhail
  Yurochkin}.} \bibinfo{year}{2021}\natexlab{}.
\newblock \showarticletitle{Post-processing for individual fairness}.
\newblock \bibinfo{journal}{\emph{Advances in Neural Information Processing
  Systems}}  \bibinfo{volume}{34} (\bibinfo{year}{2021}),
  \bibinfo{pages}{25944--25955}.
\newblock


\bibitem[\protect\citeauthoryear{Powers}{Powers}{2020}]%
        {powers2020evaluation}
\bibfield{author}{\bibinfo{person}{David~MW Powers}.}
  \bibinfo{year}{2020}\natexlab{}.
\newblock \showarticletitle{Evaluation: from precision, recall and F-measure to
  ROC, informedness, markedness and correlation}.
\newblock \bibinfo{journal}{\emph{arXiv preprint arXiv:2010.16061}}
  (\bibinfo{year}{2020}).
\newblock


\bibitem[\protect\citeauthoryear{Prost, Thain, and Bolukbasi}{Prost
  et~al\mbox{.}}{2019}]%
        {prost2019debiasing}
\bibfield{author}{\bibinfo{person}{Flavien Prost}, \bibinfo{person}{Nithum
  Thain}, {and} \bibinfo{person}{Tolga Bolukbasi}.}
  \bibinfo{year}{2019}\natexlab{}.
\newblock \showarticletitle{Debiasing embeddings for reduced gender bias in
  text classification}.
\newblock \bibinfo{journal}{\emph{arXiv preprint arXiv:1908.02810}}
  (\bibinfo{year}{2019}).
\newblock


\bibitem[\protect\citeauthoryear{Pruksachatkun, Krishna, Dhamala, Gupta, and
  Chang}{Pruksachatkun et~al\mbox{.}}{2021}]%
        {Pruksachatkun2021}
\bibfield{author}{\bibinfo{person}{Yada Pruksachatkun},
  \bibinfo{person}{Satyapriya Krishna}, \bibinfo{person}{Jwala Dhamala},
  \bibinfo{person}{Rahul Gupta}, {and} \bibinfo{person}{Kai-Wei Chang}.}
  \bibinfo{year}{2021}\natexlab{}.
\newblock \showarticletitle{Does robustness improve fairness? Approaching
  fairness with word substitution robustness methods for text classification}.
  In \bibinfo{booktitle}{\emph{ACL-IJCNLP 2021}}.
\newblock
\urldef\tempurl%
\url{https://www.amazon.science/publications/does-robustness-improve-fairness-approaching-fairness-with-word-substitution-robustness-methods-for-text-classification}
\showURL{%
\tempurl}


\bibitem[\protect\citeauthoryear{R{\"a}z}{R{\"a}z}{2022}]%
        {raz2022gerrymandering}
\bibfield{author}{\bibinfo{person}{Tim R{\"a}z}.}
  \bibinfo{year}{2022}\natexlab{}.
\newblock \showarticletitle{Gerrymandering Individual Fairness}.
\newblock \bibinfo{journal}{\emph{arXiv preprint arXiv:2204.11615}}
  (\bibinfo{year}{2022}).
\newblock


\bibitem[\protect\citeauthoryear{Richardson, Garcia-Gathright, Way, Thom, and
  Cramer}{Richardson et~al\mbox{.}}{2021}]%
        {richardson2021towards}
\bibfield{author}{\bibinfo{person}{Brianna Richardson}, \bibinfo{person}{Jean
  Garcia-Gathright}, \bibinfo{person}{Samuel~F Way}, \bibinfo{person}{Jennifer
  Thom}, {and} \bibinfo{person}{Henriette Cramer}.}
  \bibinfo{year}{2021}\natexlab{}.
\newblock \showarticletitle{Towards Fairness in Practice: A
  Practitioner-Oriented Rubric for Evaluating Fair ML Toolkits}. In
  \bibinfo{booktitle}{\emph{Proceedings of the 2021 CHI Conference on Human
  Factors in Computing Systems}}. \bibinfo{pages}{1--13}.
\newblock


\bibitem[\protect\citeauthoryear{Saleiro, Kuester, Hinkson, London, Stevens,
  Anisfeld, Rodolfa, and Ghani}{Saleiro et~al\mbox{.}}{2018}]%
        {saleiro2018aequitas}
\bibfield{author}{\bibinfo{person}{Pedro Saleiro}, \bibinfo{person}{Benedict
  Kuester}, \bibinfo{person}{Loren Hinkson}, \bibinfo{person}{Jesse London},
  \bibinfo{person}{Abby Stevens}, \bibinfo{person}{Ari Anisfeld},
  \bibinfo{person}{Kit~T Rodolfa}, {and} \bibinfo{person}{Rayid Ghani}.}
  \bibinfo{year}{2018}\natexlab{}.
\newblock \showarticletitle{Aequitas: A bias and fairness audit toolkit}.
\newblock \bibinfo{journal}{\emph{arXiv preprint arXiv:1811.05577}}
  (\bibinfo{year}{2018}).
\newblock


\bibitem[\protect\citeauthoryear{Saxena, Huang, DeFilippis, Radanovic, Parkes,
  and Liu}{Saxena et~al\mbox{.}}{2019}]%
        {saxena2019fairness}
\bibfield{author}{\bibinfo{person}{Nripsuta~Ani Saxena}, \bibinfo{person}{Karen
  Huang}, \bibinfo{person}{Evan DeFilippis}, \bibinfo{person}{Goran Radanovic},
  \bibinfo{person}{David~C Parkes}, {and} \bibinfo{person}{Yang Liu}.}
  \bibinfo{year}{2019}\natexlab{}.
\newblock \showarticletitle{How do fairness definitions fare? Examining public
  attitudes towards algorithmic definitions of fairness}. In
  \bibinfo{booktitle}{\emph{Proceedings of the 2019 AAAI/ACM Conference on AI,
  Ethics, and Society}}. \bibinfo{pages}{99--106}.
\newblock


\bibitem[\protect\citeauthoryear{Sheng, Chang, Natarajan, and Peng}{Sheng
  et~al\mbox{.}}{2019}]%
        {sheng2019woman}
\bibfield{author}{\bibinfo{person}{Emily Sheng}, \bibinfo{person}{Kai-Wei
  Chang}, \bibinfo{person}{Premkumar Natarajan}, {and} \bibinfo{person}{Nanyun
  Peng}.} \bibinfo{year}{2019}\natexlab{}.
\newblock \showarticletitle{The woman worked as a babysitter: On biases in
  language generation}.
\newblock \bibinfo{journal}{\emph{arXiv preprint arXiv:1909.01326}}
  (\bibinfo{year}{2019}).
\newblock


\bibitem[\protect\citeauthoryear{Soares, Wei, Ramamurthy, Singh, and
  Yurochkin}{Soares et~al\mbox{.}}{2022}]%
        {soares2022your}
\bibfield{author}{\bibinfo{person}{Ioana~Baldini Soares},
  \bibinfo{person}{Dennis Wei}, \bibinfo{person}{Karthikeyan~Natesan
  Ramamurthy}, \bibinfo{person}{Moninder Singh}, {and} \bibinfo{person}{Mikhail
  Yurochkin}.} \bibinfo{year}{2022}\natexlab{}.
\newblock \showarticletitle{Your Fairness May Vary: Pretrained Language Model
  Fairness in Toxic Text Classification}. In \bibinfo{booktitle}{\emph{Annual
  Meeting of the Association for Computational Linguistics}}.
\newblock


\bibitem[\protect\citeauthoryear{Srivastava, Heidari, and Krause}{Srivastava
  et~al\mbox{.}}{2019}]%
        {srivastava2019mathematical}
\bibfield{author}{\bibinfo{person}{Megha Srivastava}, \bibinfo{person}{Hoda
  Heidari}, {and} \bibinfo{person}{Andreas Krause}.}
  \bibinfo{year}{2019}\natexlab{}.
\newblock \showarticletitle{Mathematical notions vs. human perception of
  fairness: A descriptive approach to fairness for machine learning}. In
  \bibinfo{booktitle}{\emph{Proceedings of the 25th ACM SIGKDD international
  conference on knowledge discovery \& data mining}}.
  \bibinfo{pages}{2459--2468}.
\newblock


\bibitem[\protect\citeauthoryear{Stanovsky, Smith, and Zettlemoyer}{Stanovsky
  et~al\mbox{.}}{2019}]%
        {stanovsky2019evaluating}
\bibfield{author}{\bibinfo{person}{Gabriel Stanovsky}, \bibinfo{person}{Noah~A
  Smith}, {and} \bibinfo{person}{Luke Zettlemoyer}.}
  \bibinfo{year}{2019}\natexlab{}.
\newblock \showarticletitle{Evaluating gender bias in machine translation}.
\newblock \bibinfo{journal}{\emph{arXiv preprint arXiv:1906.00591}}
  (\bibinfo{year}{2019}).
\newblock


\bibitem[\protect\citeauthoryear{Veale, Van~Kleek, and Binns}{Veale
  et~al\mbox{.}}{2018}]%
        {veale2018fairness}
\bibfield{author}{\bibinfo{person}{Michael Veale}, \bibinfo{person}{Max
  Van~Kleek}, {and} \bibinfo{person}{Reuben Binns}.}
  \bibinfo{year}{2018}\natexlab{}.
\newblock \showarticletitle{Fairness and accountability design needs for
  algorithmic support in high-stakes public sector decision-making}. In
  \bibinfo{booktitle}{\emph{Proceedings of the 2018 chi conference on human
  factors in computing systems}}. \bibinfo{pages}{1--14}.
\newblock


\bibitem[\protect\citeauthoryear{Vorobyev and Krivitskaya}{Vorobyev and
  Krivitskaya}{2022}]%
        {vorobyev2022reducing}
\bibfield{author}{\bibinfo{person}{Ivan Vorobyev} {and} \bibinfo{person}{Anna
  Krivitskaya}.} \bibinfo{year}{2022}\natexlab{}.
\newblock \showarticletitle{Reducing false positives in bank anti-fraud systems
  based on rule induction in distributed tree-based models}.
\newblock \bibinfo{journal}{\emph{Computers \& Security}}
  \bibinfo{volume}{120} (\bibinfo{year}{2022}), \bibinfo{pages}{102786}.
\newblock


\bibitem[\protect\citeauthoryear{Wolf, Debut, Sanh, Chaumond, Delangue, Moi,
  Cistac, Rault, Louf, Funtowicz, et~al\mbox{.}}{Wolf et~al\mbox{.}}{2019}]%
        {wolf2019huggingface}
\bibfield{author}{\bibinfo{person}{Thomas Wolf}, \bibinfo{person}{Lysandre
  Debut}, \bibinfo{person}{Victor Sanh}, \bibinfo{person}{Julien Chaumond},
  \bibinfo{person}{Clement Delangue}, \bibinfo{person}{Anthony Moi},
  \bibinfo{person}{Pierric Cistac}, \bibinfo{person}{Tim Rault},
  \bibinfo{person}{R{\'e}mi Louf}, \bibinfo{person}{Morgan Funtowicz},
  {et~al\mbox{.}}} \bibinfo{year}{2019}\natexlab{}.
\newblock \showarticletitle{Huggingface's transformers: State-of-the-art
  natural language processing}.
\newblock \bibinfo{journal}{\emph{arXiv preprint arXiv:1910.03771}}
  (\bibinfo{year}{2019}).
\newblock


\bibitem[\protect\citeauthoryear{Woodruff, Fox, Rousso-Schindler, and
  Warshaw}{Woodruff et~al\mbox{.}}{2018}]%
        {woodruff2018qualitative}
\bibfield{author}{\bibinfo{person}{Allison Woodruff}, \bibinfo{person}{Sarah~E
  Fox}, \bibinfo{person}{Steven Rousso-Schindler}, {and}
  \bibinfo{person}{Jeffrey Warshaw}.} \bibinfo{year}{2018}\natexlab{}.
\newblock \showarticletitle{A qualitative exploration of perceptions of
  algorithmic fairness}. In \bibinfo{booktitle}{\emph{Proceedings of the 2018
  chi conference on human factors in computing systems}}.
  \bibinfo{pages}{1--14}.
\newblock


\bibitem[\protect\citeauthoryear{Xue, Yurochkin, and Sun}{Xue
  et~al\mbox{.}}{2020}]%
        {xue2020auditing}
\bibfield{author}{\bibinfo{person}{Songkai Xue}, \bibinfo{person}{Mikhail
  Yurochkin}, {and} \bibinfo{person}{Yuekai Sun}.}
  \bibinfo{year}{2020}\natexlab{}.
\newblock \showarticletitle{Auditing ml models for individual bias and
  unfairness}. In \bibinfo{booktitle}{\emph{AI\&STAT}}.
\newblock


\bibitem[\protect\citeauthoryear{Yao and Huang}{Yao and Huang}{2017}]%
        {yao2017beyond}
\bibfield{author}{\bibinfo{person}{Sirui Yao} {and} \bibinfo{person}{Bert
  Huang}.} \bibinfo{year}{2017}\natexlab{}.
\newblock \showarticletitle{Beyond parity: Fairness objectives for
  collaborative filtering}.
\newblock \bibinfo{journal}{\emph{Advances in neural information processing
  systems}}  \bibinfo{volume}{30} (\bibinfo{year}{2017}).
\newblock


\bibitem[\protect\citeauthoryear{Yurochkin, Bower, and Sun}{Yurochkin
  et~al\mbox{.}}{2020}]%
        {yurochkin2020sensr}
\bibfield{author}{\bibinfo{person}{Mikhail Yurochkin}, \bibinfo{person}{Amanda
  Bower}, {and} \bibinfo{person}{Yuekai Sun}.} \bibinfo{year}{2020}\natexlab{}.
\newblock \showarticletitle{{Training individually fair ML models with
  sensitive subspace robustness}}. In \bibinfo{booktitle}{\emph{International
  Conference on Learning Representations}}.
\newblock


\bibitem[\protect\citeauthoryear{Yurochkin and Sun}{Yurochkin and Sun}{2021}]%
        {yurochkin2021sensei}
\bibfield{author}{\bibinfo{person}{Mikhail Yurochkin} {and}
  \bibinfo{person}{Yuekai Sun}.} \bibinfo{year}{2021}\natexlab{}.
\newblock \showarticletitle{{SenSeI: Sensitive set invariance for enforcing
  individual fairness}}. In \bibinfo{booktitle}{\emph{International Conference
  on Learning Representations}}.
\newblock
\urldef\tempurl%
\url{https://github.com/IBM/inFairness}
\showURL{%
\tempurl}


\bibitem[\protect\citeauthoryear{Zotti, Carnaghi, Piccoli, and Bianchi}{Zotti
  et~al\mbox{.}}{2019}]%
        {zotti2019individual}
\bibfield{author}{\bibinfo{person}{Davide Zotti}, \bibinfo{person}{Andrea
  Carnaghi}, \bibinfo{person}{Valentina Piccoli}, {and} \bibinfo{person}{Mauro
  Bianchi}.} \bibinfo{year}{2019}\natexlab{}.
\newblock \showarticletitle{Individual and contextual factors associated with
  school staff responses to homophobic bullying}.
\newblock \bibinfo{journal}{\emph{Sexuality research and social policy}}
  \bibinfo{volume}{16}, \bibinfo{number}{4} (\bibinfo{year}{2019}),
  \bibinfo{pages}{543--558}.
\newblock


\bibitem[\protect\citeauthoryear{Zou and Schiebinger}{Zou and
  Schiebinger}{2018}]%
        {zou2018ai}
\bibfield{author}{\bibinfo{person}{James Zou} {and} \bibinfo{person}{Londa
  Schiebinger}.} \bibinfo{year}{2018}\natexlab{}.
\newblock \bibinfo{title}{AI can be sexist and racist—it’s time to make it
  fair}.
\newblock
\newblock


\end{thebibliography}

\newpage
\appendix

\section{Dataset Details}
\label{apdx:dataset-statistics}

\subsection{Dataset Statistics}
\bl{
In table \ref{tab:dataset-overall-statistics} we describe broad statistics of the dataset used to train the three models used in this work.  Overall, there were about 1.8 million data samples with only 144,334 toxic samples (8\%), making the task imbalanced. In the dataset, there are 247,581 data samples (13.71\%) that contain one of the 50 identity terms. Of these, 11.9\% or 29,509 data samples were labeled as toxic while the rest were labeled as non-toxic. In table \ref{tbl:identity_counts}, we list the counts for toxic and non-toxic data samples for each of the 50 identity terms used in this work.
}

\begin{table}[h]
\begin{tabular}{|c|c|}
\hline
\textbf{Property}                                & \textbf{Count} \\ \hline
\# of training samples                           & 1,804,874      \\ \hline
\# of toxic data samples                         & 144,334        \\ \hline
\# of non-toxic comments                         & 1,660,540      \\ \hline
\# of data samples with identity terms           & 247,581        \\ \hline
\# of toxic data samples with identity terms     & 29,509         \\ \hline
\# of non-toxic data samples with identity terms & 2,18,072       \\ \hline
\end{tabular}

\caption{\bl{Summary statistics of the toxicity-classification dataset}}
\label{tab:dataset-overall-statistics}
\end{table}

\begin{table*}
\begin{tabular}{|c|c|c|c|c|c|c|}
\cline{1-3} \cline{5-7}
\textbf{identity term} & \textbf{toxic count} & \textbf{non-toxic count} &  & \textbf{identity term} & \textbf{toxic count} & \textbf{non-toxic count} \\ \cline{1-3} \cline{5-7} 
\textbf{african} & 433 & 3116 &  & \textbf{latinx} & 0 & 12 \\ \cline{1-3} \cline{5-7} 
\textbf{african american} & 70 & 481 &  & \textbf{lesbian} & 156 & 325 \\ \cline{1-3} \cline{5-7} 
\textbf{american} & 3893 & 34267 &  & \textbf{lgbt} & 181 & 1303 \\ \cline{1-3} \cline{5-7} 
\textbf{asian} & 191 & 1905 &  & \textbf{lgbtq} & 81 & 540 \\ \cline{1-3} \cline{5-7} 
\textbf{bisexual} & 22 & 88 &  & \textbf{male} & 1049 & 6924 \\ \cline{1-3} \cline{5-7} 
\textbf{black} & 4470 & 14830 &  & \textbf{mexican} & 287 & 2124 \\ \cline{1-3} \cline{5-7} 
\textbf{blind} & 741 & 4330 &  & \textbf{middle aged} & 16 & 207 \\ \cline{1-3} \cline{5-7} 
\textbf{buddhist} & 35 & 309 &  & \textbf{middle eastern} & 44 & 394 \\ \cline{1-3} \cline{5-7} 
\textbf{canadian} & 1993 & 30979 &  & \textbf{millenial} & 7 & 69 \\ \cline{1-3} \cline{5-7} 
\textbf{catholic} & 908 & 13935 &  & \textbf{muslim} & 2274 & 8257 \\ \cline{1-3} \cline{5-7} 
\textbf{chinese} & 745 & 8337 &  & \textbf{nonbinary} & 0 & 2 \\ \cline{1-3} \cline{5-7} 
\textbf{christian} & 1284 & 8496 &  & \textbf{old} & 3778 & 35558 \\ \cline{1-3} \cline{5-7} 
\textbf{deaf} & 123 & 766 &  & \textbf{older} & 225 & 4218 \\ \cline{1-3} \cline{5-7} 
\textbf{elderly} & 119 & 2058 &  & \textbf{paralyzed} & 7 & 164 \\ \cline{1-3} \cline{5-7} 
\textbf{european} & 297 & 4251 &  & \textbf{protestant} & 42 & 1151 \\ \cline{1-3} \cline{5-7} 
\textbf{female} & 717 & 5892 &  & \textbf{queer} & 24 & 105 \\ \cline{1-3} \cline{5-7} 
\textbf{gay} & 1860 & 4390 &  & \textbf{sikh} & 23 & 265 \\ \cline{1-3} \cline{5-7} 
\textbf{heterosexual} & 127 & 488 &  & \textbf{straight} & 741 & 6025 \\ \cline{1-3} \cline{5-7} 
\textbf{hispanic} & 138 & 793 &  & \textbf{taoist} & 0 & 5 \\ \cline{1-3} \cline{5-7} 
\textbf{homosexual} & 368 & 999 &  & \textbf{teenage} & 90 & 637 \\ \cline{1-3} \cline{5-7} 
\textbf{indian} & 246 & 2641 &  & \textbf{trans} & 157 & 1426 \\ \cline{1-3} \cline{5-7} 
\textbf{japanese} & 176 & 2313 &  & \textbf{transgender} & 296 & 1264 \\ \cline{1-3} \cline{5-7} 
\textbf{jewish} & 403 & 3002 &  & \textbf{white} & 8641 & 29767 \\ \cline{1-3} \cline{5-7} 
\textbf{latina} & 6 & 50 &  & \textbf{young} & 1637 & 17938 \\ \cline{1-3} \cline{5-7} 
\textbf{latino} & 76 & 434 &  & \textbf{younger} & 167 & 2763 \\ \cline{1-3} \cline{5-7} 
\end{tabular}

    \caption{\bl{Number of data samples containing the corresponding identity term labeled as toxic and non-toxic examples.}}
    \label{tbl:identity_counts}
\end{table*}

\subsection{Toxicity Annotation Details}

\bl{
    To label the data samples as toxic or non-toxic, dataset curators used human annotators to label comments as 0 (non-toxic) and 1 (toxic). 
    Multiple annotators annotated each data sample and the mean of their assigned labels was recorded as the final toxicity confidence score. Following prior work \citep{garg2019counterfactual,yurochkin2021sensei,koh2021wilds}, we marked data samples with greater than or equal to 0.5 toxicity confidence score as toxic and others as non-toxic. In Table \ref{table:annotator-stats}, we list the various quantiles of annotator counts for data samples, along with the mean toxicity score in the dataset. Figure \ref{fig:annotator-confidence-hist} provides a histogram of the mean annotator confidence scores. We note that annotators demonstrated a lot more agreement when labeling comments as non-toxic.
}

\begin{table*}
\begin{minipage}[b]{0.45\linewidth}
\centering
\begin{tabular}{|c|c|}
    \hline
    \textbf{Property} & \textbf{Count} \\ \hline
    Minimum annotator count & 3 \\ \hline
    Median annotator count & 4 \\ \hline
    Mean annotator count & 8.78 \\ \hline
    Maximum annotator count & 4936 \\ \hline
    Mean (std) toxicity score & 0.103 $\pm$ 0.197 \\ \hline
    \end{tabular}
    \caption{
        \bl{Various quantiles of the number of annotators marking a sample as toxic or not in the dataset. Also mentioned is the mean toxicity score across all data samples in the dataset.}
    }
    \label{table:annotator-stats}
\end{minipage}\hfill
\begin{minipage}[b]{0.45\linewidth}
\centering
\includegraphics[width=0.8\textwidth]{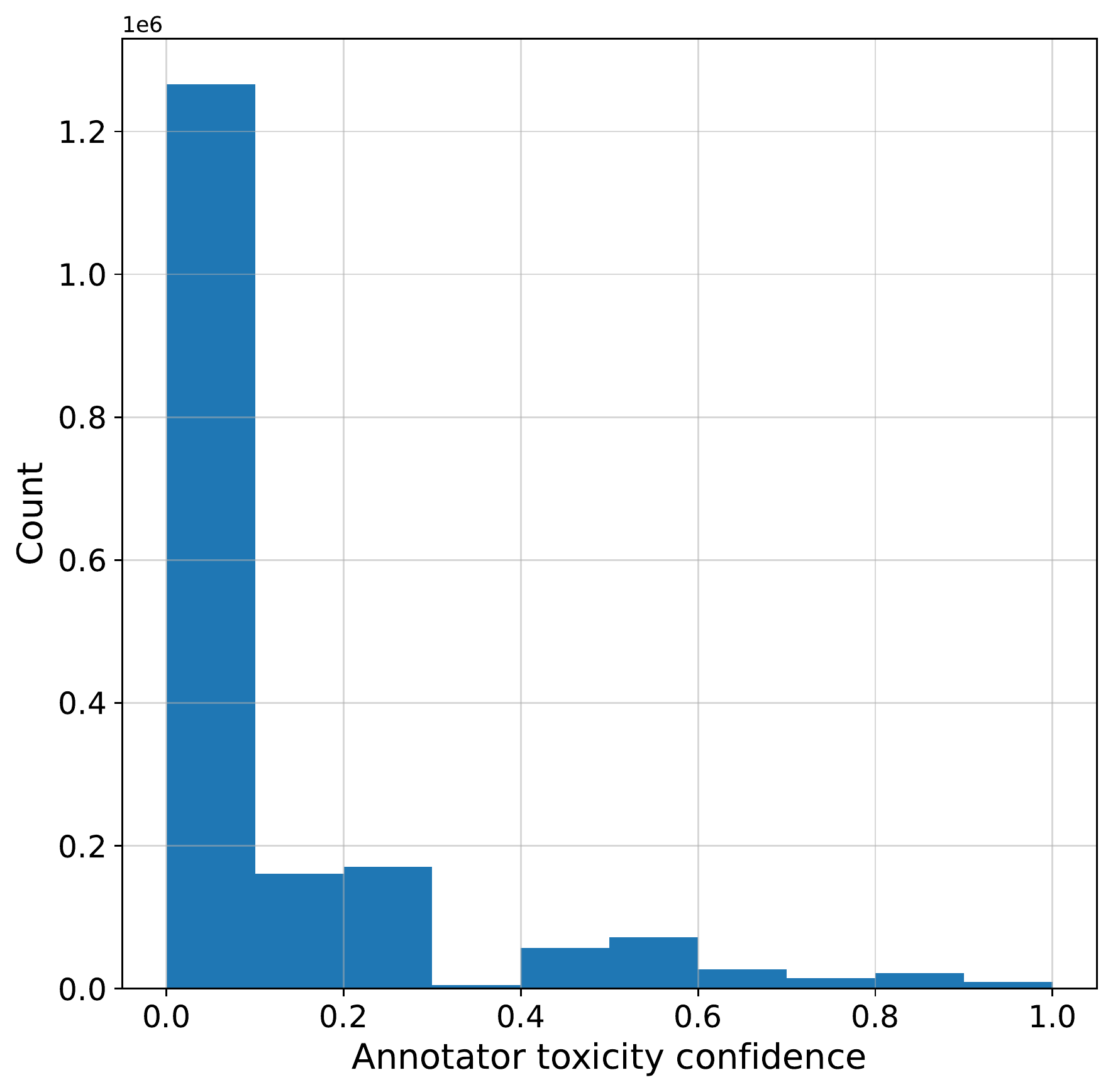}
\captionof{figure}{
\bl{Histogram of annotator confidence score in data samples being toxic or non-toxic.}
}
\label{fig:annotator-confidence-hist}
\end{minipage}
\end{table*}

\section{Model training details}
\label{apdx:model-training}

\begin{figure*}
    \centering
    \includegraphics[width=\textwidth]{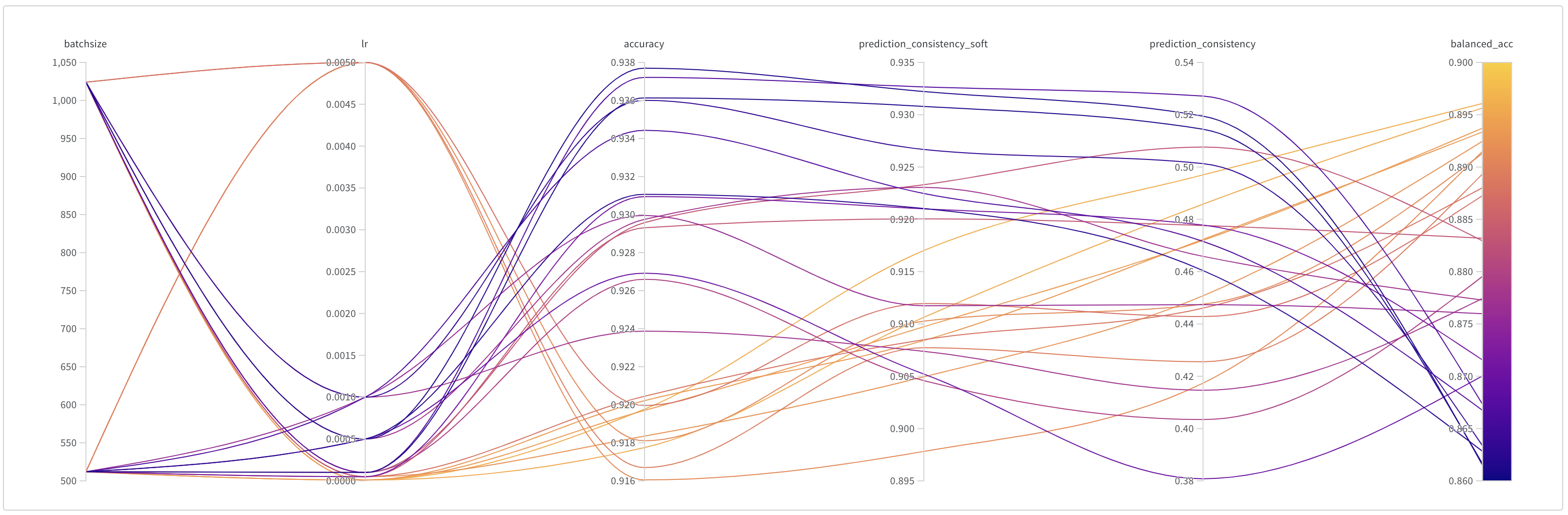}
    \caption{\bl{Hyperparameter sweep ranges for training Model A. We vary the batch size and learning rate values for Model A and measure their accuracy, soft prediction consistency, prediction consistency, and balanced accuracy. We select the final model based on its performance on prediction consistency and balanced accuracy.}}
    \label{fig:parallel-coords-standard}
\end{figure*}

\begin{figure*}
    \centering
    \includegraphics[width=\textwidth]{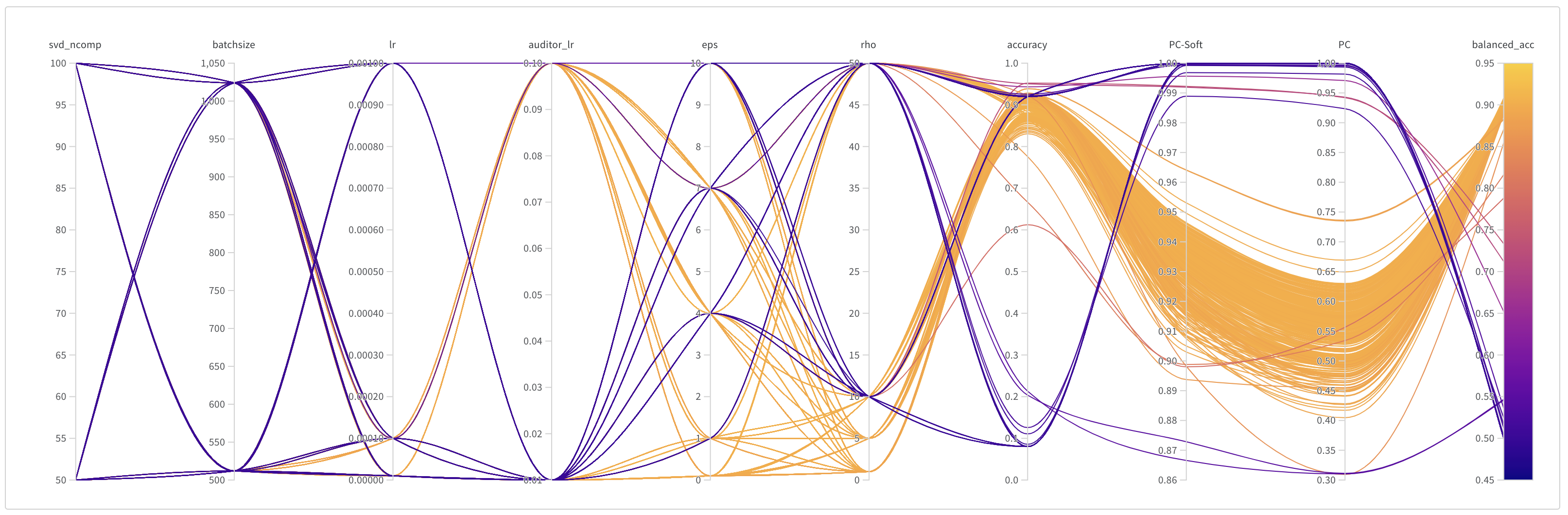}
    \caption{\bl{Hyperparameter sweep ranges for training Model B. We vary six hyperparameters for Model B (described in Table \ref{tbl:hyperparameters}) and measure their accuracy, soft prediction consistency, prediction consistency, and balanced accuracy. We select the final model based on its performance on prediction consistency and balanced accuracy.}}
    \label{fig:parallel-coords-fair}
\end{figure*}

\bl{To train Model A (standard model) and Model B (individually fair model), we optimized hyperparameters over a grid of values defined in Table \ref{tbl:hyperparameters}. For each hyperparameter setting, we compute the Accuracy, Balanced Accuracy, and Prediction Consistency on the validation set, and selected a final model with good Balanced Accuracy and Prediction Consistency scores. We display the performance of the various models trained with different hyperparameter values in the parallel coordinates Figure \ref{fig:parallel-coords-standard} and Figure \ref{fig:parallel-coords-fair}. To train Model B (group fair model) we experimented with multiple types of constraints on the performance of each of the groups (false positive rate, false negative rate, and balanced accuracy) and identified that constraining balanced accuracy of each group to be $\geq 95\%$ worked best as mentioned in Section \ref{sec:models:group}. Additionally, we optimized learning rate across the grid of values defined in Table \ref{tbl:hyperparameters}.}

\begin{table*}
    \centering
    \begin{tabular}{|c|c|c|c|}
        \hline
        \textbf{Hyperparameters} & \textbf{Model A}                       & \textbf{Model B}         & \textbf{Model C}    \\ \hline
        Batch size               & \{512, 1024\}                          & \{512, 1024\}            &   \{1024\} \\ \hline
        learning rate            & \{5e-5, 1e-5, 5e-4, 1e-4, 5e-3, 1e-3\} & \{1e-5, 1e-4, 1e-3\}   & \{0.03, 0.035, 0.04\}      \\ \hline
        rho                      & N.A.                                   & \{1.0, 5.0, 10.0, 50.0\}    &   N.A.    \\ \hline
        eps                      & N.A.                                    & \{0.1, 1.0, 4.0, 7.0, 10.0\}        &   N.A.    \\ \hline
        auditor learning rate    & N.A.                                    & \{0.1, 0.01\}               &   N.A.            \\ \hline
        SVD num components       & N.A.                                   & \{50, 100\}                 &   N.A.            \\ \hline
    \end{tabular}
        
    \caption{\bl{Hyperparameter grid-search ranges for Models A, B, and C. We evaluate the trained models for their balanced accuracy and prediction consistency and select the model with good performance on both metrics.}}
    \label{tbl:hyperparameters}
\end{table*}

\end{document}